\documentclass{article}
\usepackage[utf8]{inputenc}
\usepackage{bm,fullpage}
\usepackage{amsmath, amssymb, amsfonts,amsthm}
\usepackage{graphicx}
\usepackage{epstopdf}
\usepackage{booktabs}
\usepackage{float}
\usepackage{subfigure}
\usepackage{natbib}
\usepackage{caption}
\newtheorem{theorem}{Theorem}
\newtheorem{lemma}{Lemma}
\newtheorem{remark}{Remark}
\newtheorem{assumption}{Assumption}
\newtheorem{corollary}{Corollary}
\usepackage{enumerate}
\def\cp{\mathop{\rightarrow}\limits^{p}}

\def\I{\mathbf{I}}
\allowdisplaybreaks[4]
\title{Inverse Norm Weighted Maxsum Test for High Dimensional Location Parameters}
\author{Guowei Yan, Ping Zhao and Long Feng \\
School of Statistics and Data Science, LMPC and LEBPS, Nankai University}
\date{}

\begin{document}

\maketitle

\begin{abstract}
In the context of high-dimensional data, we investigate the one-sample location testing problem. We introduce a max-type test based on the weighted spatial sign, which exhibits exceptional performance, particularly in the presence of sparse alternatives. Notably, we find that the inverse norm test significantly enhances the power of the test compared to several existing max-type tests. Next, we prove the asymptotic independence between the newly proposed max-type test statistic and the sum-type test statistic based on the weighted spatial sign. Then, we propose an innovative max-sum type testing procedure that integrates both test statistics. This novel procedure demonstrates remarkable robustness and effectiveness across a wide range of signal sparsity levels and heavy-tailed distributions. Through extensive simulation studies, we highlight the superior performance of the proposed method, showcasing its robustness and efficiency compared to traditional alternatives in various high-dimensional settings.

\end{abstract}
\section{Introduction}

In multivariate statistics, testing location parameters plays a central role. When the number of variables, $ p $, is smaller than the sample size, $ n $, Hotelling's $ T^2 $ test is highly effective. However, as we transition to high-dimensional data, where $ p $ exceeds $ n $, practical challenges arise. The primary issue stems from the dependence of these tests on the invertibility of the sample covariance matrix. As a result, there is a critical need to develop novel methodologies capable of efficiently addressing mean testing in high-dimensional settings.

In high-dimensional settings, a straightforward approach often involves replacing the Mahalanobis distance with the Euclidean distance. For two-sample problems, \cite{bai} introduced an $L^2$-norm-based statistic that considers the Euclidean distance between sample means. This method was later refined by \cite{CQ}, who improved its efficiency by eliminating redundant terms, resulting in a test statistic that is independent of the dimension-sample size relationship. Further advancements were made by \cite{Sri}, \cite{PA}, and \cite{F15}, all of which led to scalar-invariant test statistics. These statistics replace the sample covariance matrix with its diagonal form. However, the above methods typically assume normality or diverging factor models. In contrast, many real-world datasets exhibit heavy-tailed characteristics that violate these assumptions. For instance, multivariate $t$-distributions often fail to satisfy the underlying assumptions of the previously mentioned methods. In response to these challenges, robust testing procedures have been developed to address such distributions, ensuring accurate results even in the presence of heavy-tailed data.

The extension of classical univariate rank and signed-rank methods to the multivariate setting has proven effective for high-dimensional data analysis. Spatial sign methods, such as those introduced by \cite{W15}, \cite{FS}, and \cite{F16}, are notable for their incorporation of the Oja median \citep{Oja}. However, these methods primarily focus on the direction of observations, neglecting the information contained in their magnitude. \cite{FLM21} propose a novel inverse norm sign test (INST) that is consistent and demonstrates significantly greater power compared to the aforementioned popular tests. Furthermore, they introduce a general class of weighted spatial sign tests that encompasses these existing methods, proving that INST is the optimal test within this class, being both consistent and uniformly more powerful than all other members. \cite{H23} further extend this approach to high-dimensional two-sample problems.

All of the aforementioned methods are sum-type test procedures, which rely on the \(L^2\)-norm of the sample mean or spatial median. These methods perform well under dense alternatives, where a large number of variables exhibit non-zero means. However, they face challenges when applied to sparse alternatives. To address this issue, max-type test procedures have been developed. \cite{cai} introduced a test statistic based on the maximum difference between sample means, assuming Gaussian or sub-Gaussian distributions. For heavy-tailed distributions, \cite{C23} applied a Gaussian approximation along with a multiplier bootstrap algorithm to the sample spatial median. However, this approach lacks scalar invariance and is computationally intensive. \cite{L24} proposed a spatial-sign-based max-type test statistic, which demonstrates superior performance for sparse alternatives.

Motivated by the work of \cite{FLM21} and \cite{H23}, we propose a weighted spatial-sign-based max-type test procedure that incorporates both the direction and the magnitude (modulus) of the observations, and includes the aforementioned max-type tests as special cases. We establish the asymptotic distribution of the corresponding test statistics and show that the optimal weighted function remains \(1/t\). Theoretical results and simulation studies indicate that the proposed optimal max-type test procedure outperforms existing max-type tests under sparse alternatives and heavy-tailed distributions.

In practical applications, the nature of the alternatives—whether dense or sparse—is often unknown, which has led to the development of adaptive strategies that combine sum-type and max-type tests. For instance, \cite{X16} incorporated different \(L_n\)-norms of sample means. Additionally, \cite{H21} proposed a family of U-statistics as unbiased estimators of the \(L_n\)-norms of mean vectors. These U-statistics exhibit asymptotic independence and asymptotically follow a normal distribution, distinguishing them from max-type test statistics. Further research by \cite{F22a} demonstrated the independence between sum-type and max-type statistics under less restrictive covariance matrix assumptions, thus enhancing their applicability in a wide range of high-dimensional problems. Recent studies have focused on the asymptotic independence between sum-type and max-type statistics in the context of heavy-tailed distributions. In particular, \cite{L24} established the asymptotic independence between the max-type test statistic and the spatial-sign-based sum-type test statistic proposed by \cite{FS}.

In this paper, we first introduce a newly proposed max-type test statistic and demonstrate that it is asymptotically independent of the existing sum-type test statistic, which is based on the weighted spatial-sign method proposed by \cite{FLM21}. To combine these two test statistics effectively, we construct a Cauchy combination test procedure. Theoretical results, as well as extensive simulation studies, show that the proposed Cauchy combination test exhibits excellent performance across a broad spectrum of sparsity levels of the alternative hypothesis and under heavy-tailed distributions. Furthermore, a real-data application illustrates the superiority of the newly proposed method over existing approaches, highlighting its robustness and versatility in practical settings. These results suggest that the Cauchy combination test provides a powerful and reliable tool for hypothesis testing in challenging scenarios where traditional methods may fall short.

The paper is organized as follows. In Section 2, we derive the Bahadur representation of the weighted spatial median and formulate the corresponding max-type test statistic. Section 3 is devoted to recalling the weighted sum-type test statistic and rigorously proving the asymptotic independence between the two statistics. In Section 4, we present the results from a series of simulation studies, which illustrate the performance of the proposed methods under various scenarios. Section 5 applies the proposed approach to a real-data example, demonstrating its practical utility and superiority over existing methods. All technical details, including proofs and supplementary material, are provided in the Appendix for reference.

{\it Notations}: For $d$-dimensional $\bm{x}$, we use the notation $\|\bm{x}\|$ and $\|\bm{x}\|_\infty$ to denote its Euclidean norm and maximum-norm respectively. The spatial sign function is defined as $U(\bm{x})=\|\bm{x}\|^{-1}\bm{x}\bm{1}(\bm{x}\neq \bm{0})$. Denote $a_n\lesssim b_n$ if there exists constant $C$, $a_{n}\leq$ $Cb_{n}$ and $a_n\asymp b_n$ if both $a_n\lesssim b_n$ and $b_n\lesssim a_n$ hold. Let $\psi_\alpha(x)=\exp\left(x^{\alpha}\right)-1$ be a function defined on $[0,\infty)$ for $\alpha>0.$ Then the Orlicz norm $\|\cdot\|_{\psi_{\alpha}}$ of a $\boldsymbol{X}$ is defined as $\|\boldsymbol{X}\|_{\psi_\alpha}=\inf\left\{t>0,\mathbb{E}\left\{\psi_\alpha(|\boldsymbol{X}|/t)\right\}\leqslant1\right\}.\:$Let $\mathrm{tr}(\cdot)$ be a trace for matrix, $\lambda_{min}(\cdot)$ and $\lambda_{max}(\cdot)$ be the minimum and maximum eigenvalue for symmetric martix. $\mathbf{I}_p$ represents a p-dimensional identity matrix, and diag$\{v_1,v_2,\cdots,v_p\}$ represents the diagonal matrix with entries $\boldsymbol{v}= ( v_1, v_2, \cdots , v_p)$. For $a, b\in \mathbb{R}$, we write $a\land b= \min \{ a, b\}$.

\section{Max-type test}
Let $\bm{X}_1,\ldots,\bm{X}_n$ be a sequence of independent and identically distributed (i.i.d.) $p$-dimensional random vectors drawn from a population $\bm{X}$ having a cumulative distribution function $F_X$ in $\mathbb{R}^p$. For each index $i$, the random vector $\bm{X}_i$ can be expressed in the following form:
\begin{equation}\label{model}
    \bm{X}_i=\bm{\theta}+v_i\mathbf{\Gamma}\bm{W}_i,
\end{equation}
where $\bm{\theta}$ represents the location parameter. The vector $\bm{W}_i$ is a $p$-dimensional random vector with independent components, satisfying $\mathbb{E}(\bm{W}_i)=\bm{0}$. The covariance matrix $\mathbf{\Sigma}$ is given by $\mathbf{\Sigma}=\mathbf{\Gamma}\mathbf{\Gamma}^\top$. Additionally, $v_i$ is a non-negative univariate random variable that is independent of the spatial sign of $\bm{W}_i$.

In this paper, our focus is on the following hypothesis-testing problem:
\begin{equation}
    H_0:\bm{\theta}=\bm{0} \quad \text{versus} \quad H_1: \bm{\theta}\neq \bm{0}.
\end{equation}
It is widely recognized that max-type tests exhibit good performance under sparse alternatives, as noted in \cite{cai}. The spatial sign function is defined as $U(\bm{x})=\|\bm{x}\|\bm{x}\bm{I}(\bm{x}\neq\bm{0})$. In \cite{C23}, a maximum-norm type test statistic has been proposed:
\begin{equation*}
    T_{CPZ}=n^{1/2}\|\check{\bm{\theta}}_n\|_{\infty},
\end{equation*}
where $\check{\bm{\theta}}_n$ is such that $\sum_{i = 1}^n U(\bm{X}_i-\check{\bm{\theta}}_n)=0$ holds. Under the null hypothesis $H_0$, the multiplier bootstrap method is employed to approximate the distribution of $T_{CPZ}$.

In \cite{L24}, a spatial-sign based max-type test statistic has been put forward:
\begin{equation}
    T_{SS-MAX}=n\|\hat{\mathbf{D}}^{-1/2}\hat{\bm{\theta}}_n\|_{\infty}^2\hat{\zeta}_{-1}^2p(1-n^{-1/2}).
\end{equation}
Here, $\hat{\zeta}_{-1}:=\frac{1}{n}\sum_{i = 1}^n\left\|\hat{\mathbf{D}}^{-1/2}\left( \bm{X}_i-\hat{\bm{\theta}}_n \right)\right\|^{-1}$, and $\hat{\bm{\theta}}_n$ and $\hat{\mathbf{D}}$ are the estimators of the location parameter and the diagonal matrix of the covariance matrix $\mathbf{\Sigma}$, respectively. These estimators satisfy the following equations:
\begin{equation}\label{oldm}
\begin{aligned}
    &\frac{1}{n}\sum_{i = 1}^n U\left(\mathbf{\hat{D}}^{-1/2}\left( \bm{X}_i-\bm{\hat{\theta}}_n \right)\right)=0;\\
    &\frac{p}{n}\;\mathrm{diag}\left\{\sum_{i = 1}^nU\left(\mathbf{\hat{D}}^{-1/2}\left( \bm{X}_i-\bm{\hat{\theta}}_n\right)\right)U\left(\mathbf{\hat{D}}^{-1/2}\left( \bm{X}_i-\bm{\hat{\theta}}_n \right)\right)^{\top}\right\}=\mathbf{I}_p.
\end{aligned}
\end{equation}
The work in \cite{L24} provides the Bahadur representation and Gaussian approximation of the location parameter. By utilizing the estimators of the diagonal matrix of the covariance matrix, $T_{SS-MAX}$ is scalar-invariant and attains better power performance.
\subsection{Bahadur representation of weighted max-type test}
As demonstrated in \cite{FLM21} and \cite{H23}, incorporating weighted spatial-sign methods that account for the magnitude of the observations can significantly improve power performance. So, motivated by (\ref{oldm}), we propose a weighted approach to find a pair of diagonal matrix $\mathbf{D}$ and vector $\bm{\theta}$ for each sample that simultaneously satisfy
\begin{equation}\label{newm}
\begin{aligned}
    &\frac{1}{n}\sum_{i=1}^n w\left(\left\|\mathbf{{D}}^{-1/2}\left( \bm{X}_i-\bm{{\theta}} \right)\right\|\right)U\left(\mathbf{{D}}^{-1/2}\left( \bm{X}_i-\bm{{\theta}} \right)\right)=0;\\
    &\frac{p}{n}\;\mathrm{diag}\left\{\sum_{i=1}^nU\left(\mathbf{{D}}^{-1/2}\left( \bm{X}_i-\bm{{\theta}} \right)\right)U\left(\mathbf{{D}}^{-1/2}\left( \bm{X}_i-\bm{{\theta}} \right)\right)^{\top}\right\}=\mathbf{I}_p.
\end{aligned}
\end{equation}
where $w(\cdot)$ is a continuous nonnegative weight function. $\left(\bm{\theta}, \mathbf{D} \right)$ can be viewed as a simplified version of HettmanspergerRandles(HR) estimator \cite{HR} without considering the off-diagonal elements of $\mathbf{S}$. We could solve Equation (\ref{newm}) by the following three steps until convergence:
\begin{align*}
    &(i)\ \bm\epsilon_i\leftarrow\mathbf{D}^{-1/2}\left( \bm{X}_i -\bm{\theta}\right),\qquad i=1,\dots,n; \\
    &(ii)\ \bm{\theta}\leftarrow\bm{\theta}+\frac{\mathbf{D}^{-1/2}\sum_{i=1}^n w(\|\bm{\epsilon}_i\|)U(\bm{\epsilon}_i)}{\sum_{i=1}^nw(\|\bm{\epsilon}_i\|)\|\bm{\epsilon}_i\|^{-1}};\\
    &(iii)\ \mathbf{D}\leftarrow p\mathbf{D}^{1/2}\mathrm{diag}\left\{n^{-1}\sum_{i=1}^n U(\bm{\epsilon}_i)U(\bm{\epsilon}_i)^\top\right\}\mathbf{D}^{-1/2}.
\end{align*}
For convenience, the initial estimators can be set using the sample mean and sample variances. The resulting estimators of the location and the diagonal matrix are denoted as $\hat{\bm{\theta}}$ and $\hat{\mathbf{D}}$, respectively. Evidently, these two estimators are related to the weight function $w(\cdot)$. Unfortunately, similar to a common shortcoming of the HR estimate, there is no proof that the aforementioned algorithm converges, even in low-dimensional scenarios, although it is consistently effective in practice. At the same time, the existence and uniqueness of the estimators also lack proofs. Consequently, we present our first assumption.
\begin{assumption}\label{existence and uniqueness of HR}
    The Equations \ref{newm} exist an unique solution $\left(  \bm{\hat{\theta}},\hat{\mathbf{D}}\right)$.
\end{assumption}

In the remaining part of this section, we explore the Bahadur representation and Gaussian approximation of $\hat{\mathbf{D}}^{-1/2}(\hat{\bm{\theta}}-\bm{\theta})$. Leveraging these, we can deduce the limiting distribution of $\hat{\bm{\theta}}$. For $i = 1,2,\ldots,n$, let $\bm{U}_i = U\left(\mathbf{D}^{-1/2}\left( \bm{X}_i-\bm{\theta} \right)\right)$ and $R_i=\left\|\mathbf{D}^{-1/2}\left( \bm{X}_i-\bm{\theta} \right)\right\|$, which represent the scale-invariant spatial-sign and radius of $\bm{X}_i-\bm{\theta}$, respectively. The $k$-th moment of $R_i$ is defined as $\zeta_k=\mathbb{E}(R_i^k)$. Denoting $\bm{W}=(W_{i,1},\ldots, W_{i,p})^\top$, the following is the assumption.
\begin{assumption}\label{struction of W}
    $W_{i,1},\dots,W_{i,p}$ are i.i.d. symmetric random variables with $\mathbb{E}(W_{i,j})=0$, $\mathbb{E}(W_{i,j}^2)=1$, and $\|W_{i,j}\|_{\psi_\alpha}\le c_0$ with some constant $c_0>0$ and $1\le \alpha \le 2$.
\end{assumption}
\begin{assumption}\label{weighted function}
    The weighted function $w(x)=x^m$, $m\le 1$.
\end{assumption}
\begin{assumption}\label{order}
    there exist two positive constants $\underline{b}$ and $\bar{B}$ such that $\underline{b}\le \lim\sup_p \mathbb{E}(R_i/\sqrt{p})^{k}\le \bar{B}$ for $k\in \{m,2m,3m,4m,m-2,2m-4,m+1,2m+2,-1,-2,-3,-4\}$.
\end{assumption}
\begin{assumption}\label{bound of R}
    The shape matrix $\mathbf{R}=\mathbf{D}^{-1/2}\mathbf{\Sigma}\mathbf{D}^{-1/2}=(\sigma_{jl})_{p\times p}$ satisfies $\max_{j=1,\dots,p}\sum_{l=1}^p|\sigma_{jl}|\le a_{0}(p)$ In addition, $\lim \inf_{p\rightarrow\infty}\min_{j=1,\dots,p}d_j>\underline{d}$ for some constant $\underline{d}>0$, where $\mathbf{D}=\mathrm{diag}\{d_1^2,d_2^2,\dots,d_p^2\}$.
    \end{assumption}
\begin{assumption}\label{R2}
    (i) $\mathrm{tr}(\mathbf{R}^2)-p=o(n^{-1}p^2)$, (ii) $n^{-2}p^2/\mathrm{tr}(\mathbf{R}^2)=O(1)$ and $\log p=o(n)$.
\end{assumption}
\begin{remark}
    Assumption \ref{struction of W} is identical to Condition C.1 in \cite{C23}. This assumption ensures that $\bm{\theta}$ in model \ref{model} represents the population spatial median, and that $W_{i,j}$ follows a sub-exponential distribution. In the case where $W_i\sim N(\bm{0},\mathbf{I}_p)$, $\bm{X}_i$ adheres to an elliptical symmetric distribution. In Assumption \ref{weighted function}, we set $w = x^m$ with $m\leq1$. In practice, a more general weighted function can be employed. Such a function must satisfy the conditions that $w(x)>0$ for all $x$ and $w'(x)\to0$ as $x\to+\infty$. Concurrently, we would need to modify Assumption \ref{order} into a more intricate form, which would incorporate the moments of $w(R_i)$, $w'(R_i)$, and $R_i w'(R_i)$. However, in reality, we typically only use $w(x)=x$, $w(x)=1$, or $w(x)=\frac{1}{x}$. For simplicity's sake, we have restricted our consideration to $w(x)=x^m$.

\end{remark}
\begin{theorem}\label{Bahadur represengtation}
    Under Assumptions 1-6 and $a_0(p)\asymp p^{1-\delta}$, if $\log p=o(n^{1/3})$ and $\log n=o(p^{1/3\wedge\delta})$, we have that
    \begin{equation*}
        n^{1/2}\hat{\mathbf{D}}^{-1/2}(\hat{\bm{\theta}}-\bm{\theta})=n^{-1/2}\zeta_{m-1}^{-1}\sum_{i=1}^n w(R_i)\bm{U}_i+C_n,
    \end{equation*}
    where
    \begin{equation*}
        \|C_n\|_\infty=O_{p}(n^{-1/4}\log^{1/2}(np)+p^{-(1/6\wedge\delta/2)}\log^{1/2}(np)+n^{-1/2}(\log p)^{1/2}\log^{1/2}(np)).
    \end{equation*}
\end{theorem}
$\operatorname{Let}\mathcal{A}^{\operatorname{re}}=\left\{\prod_{j=1}^p\left[a_j,b_j\right]:-\infty\leqslant a_j\leqslant b_j\leqslant\infty,j=1,\ldots,p\right\}$be the class of rectangles in $\mathbb{R}^p.$ Based on the Bahadur representation of $\hat{\boldsymbol{\theta}}_n$, we acquire the following Gaussian approximation of $\hat{\mathbf{D}}^{-1/2}\left(\hat{\boldsymbol{\theta}}-\boldsymbol{\theta}\right)$ in rectangle $\mathcal{A}^\mathrm{re}.$

\begin{lemma}\label{GA}
    Under Assumptions 1-6,
    \begin{equation*}
        \rho_n\left(\mathcal{A}^\mathrm{re}\right)=\sup_{A\in \mathcal{A}^\mathrm{re}}\left|\mathbb{P}\left\{n^{1/2}\hat{\mathbf{D}}^{-1/2}\left(\hat{\boldsymbol{\theta}}-\boldsymbol{\theta}\right)\in A\right\}-\mathbb{P}(\boldsymbol{G}\in A)\right|\to0,
    \end{equation*}
    as $n\rightarrow\infty$, where $\bm{G}\sim N\left(0,\zeta_{m-1}^{-2}\mathbf{\Sigma}_{w}\right)$ with $\mathbf{\Sigma}_w=\mathbb{E}(w^2(R_1)\bm{U}_1\bm{U}_1^\top)$.
\end{lemma}
The Gaussian approximation for $\hat{\bm{\theta}}$ suggests that the probabilities $\mathbb{P}\left\{n^{1/2}\hat{\mathbf{D}}^{-1/2}\left(\hat{\bm{\theta}}-\bm{\theta}\right)\right\}$ can be approximated by those of a centered Gaussian random vector with a covariance matrix $\zeta_{m-1}^{-2}\mathbf{\Sigma}_w$ for hyperrectangles $A\in \mathcal{A}^{re}$.

Denote $\mathcal{A}^{t}=\left\{\prod_{j=1}^{p}\left[a_{j},b_{j}\right]:-\infty=a_{j}\leqslant b_{j}=t,j=1,\ldots,p\right\}$. Then, set $A\in \mathcal{A}^{t}$ we have the following corollary.
\begin{corollary}\label{GAmax}
    Suppose the conditions assumed in Lemma \ref{GA} hold, as $n\to\infty$,
    \begin{equation*}        \rho_{n}=\sup\limits_{t\in\mathbb{R}}\left|\mathbb{P}\left(n^{1/2}\|\hat{\mathbf{D}}^{-1/2}(\hat{\boldsymbol{\theta}}-\boldsymbol{\theta})\|_{\infty}\leqslant t\right)-\mathbb{P}\left(\|\boldsymbol{G}\|_{\infty}\leqslant t\right)\right|\to0,
    \end{equation*}
    where $\mathbf{G}\sim N(0,\zeta_{m-1}^{-2}\mathbf{\Sigma}_{w})$.
\end{corollary}
Moreover, we can find covariance matrix $\mathbf{\Sigma}_w$ is related to $\zeta_{2m}p^{-1}\mathbf{R}$ as $p\to\infty$. Thus we could have the following lemma.
\begin{lemma}\label{VA}
    Suppose $\bm{G}\sim N(0,\zeta_{m-1}^{-2}\mathbf{\Sigma}_w)$ and $\bm{Z}\sim N(0,\zeta_{m-1}^{-2}\zeta_{2m}p^{-1}\mathbf{R})$, as $(n,p)\to \infty$,
    \begin{equation*}        \sup_{t\in\mathbb{R}}|\mathbb{P}\left(\|\boldsymbol{Z}\|_{\infty}\leqslant t\right)-\mathbb{P}\left(\|\boldsymbol{G}\|_{\infty}\leqslant t\right)|\to0.
    \end{equation*}
\end{lemma}
Combining Corollary \ref{GAmax} and Lemma \ref{VA}, we get the following theorem of Gaussian approximation
\begin{theorem}\label{GA2}
    Under Assumptions 1-6 and $a_0(p)=p^{1-\delta}$ for some position constant $\delta\le1/2$. If $\log p=o(n^{1/5})$ and $\log n=o(p^{1/3\wedge\delta})$, then
    \begin{equation*}
    \widetilde{\rho}_n=\sup\limits_{t\in\mathbb{R}}\left|\mathbb{P}\left(n^{1/2}\|\hat{\mathbf{D}}^{-1/2}(\hat{\boldsymbol{\theta}}-\boldsymbol{\theta})\|_{\infty}\leqslant t\right)-\mathbb{P}\left(\|\boldsymbol{Z}\|_{\infty}\leqslant t\right)\right|\to0,
    \end{equation*}
    where $\bm{Z}\sim N(0,\zeta_{m-1}^{-2}\zeta_{2m}p^{-1}\mathbf{R})$.
\end{theorem}
\subsection{Weighted max-type test procedure}
At present, we possess the weighted Bahadur representation and its Gaussian approximation. To present our statistics, we introduce the following assumption.
\begin{assumption}\label{sparse of R}
    Let $\mathbf{R}=(\sigma_{ij})_{1\le i,j\le p}$. For some $\varrho\in (0,1)$. assume $|\sigma_{ij}|\le \varrho$ for all $1\le i<j\le p$ and $p\ge 2$. Suppose $\{\delta_p;p\ge 1\}$ and $\{\kappa_p;p\ge 1\}$ are positive constants with $\delta_p=o(1/\log p)$ and $\kappa=\kappa_p\rightarrow 0$ as $p\rightarrow\infty$. For $1\le i\le p$, define $B_{p,i}=\{1\le j\le p;|\sigma_{i,j}|\ge\delta_p\}$ and $C_{p}=\{1\le i\le p;|B_{p,i}|\ge p^\kappa\}$. We assume that $|C_p|/p\rightarrow 0$ as $p\rightarrow\infty$.
\end{assumption}
The fulfillment of Assumption \ref{sparse of R} is crucial, as it ensures that the maximum value of a sequence of normal variables converges to a Gumbel limiting distribution. This assumption specifically characterizes the required correlation structure among the variables to achieve the desired result.

Assume that Assumptions 1-6 hold. Then, according to Theorem 2 in \cite{F22a}, the distribution of $\|\boldsymbol{Z}\|_{\infty}^2$ in Theorem \ref{GA2} is a Gumbel distribution with cumulative distribution function (cdf) $F(x)=\exp\left\{-\frac{1}{\sqrt{\pi}}e^{-x/2}\right\}$. In other words,
\begin{equation}
    \mathbb{P}\left(n\left\|\hat{\mathbf{D}}^{1/2}(\hat{\bm{\theta}}-\bm{\theta})\right\|^2_\infty p\zeta_{m-1}^2\zeta_{2m}^{-1}-2\log p+\log\log p\leq x\right)\to \exp\left\{-\frac{1}{\sqrt{\pi}}e^{-x/2}\right\}.
\end{equation}
Consequently, it is natural for us to consider the following statistics based on the scalar-invariant spatial median $\hat{\bm{\theta}}$ in order to enhance the convergence rate of the maximum. We define
\begin{equation*}
    T_{MAX}^{(m)}=n\left\|\hat{\mathbf{D}}^{-1/2}\hat{\bm{\theta}}\right\|^2_{\infty}\hat{\zeta}_{m-1}^2\hat{\zeta}_{2m}^{-1}p\cdot(1-n^{-1/2})-2\log p+\log\log p,
\end{equation*}
where $\hat{\zeta}_k=n^{-1}\sum_{i=1}^n \|\hat{\mathbf{D}}^{-1/2}(\bm X_i-\hat{\bm \theta})\|^{k}$.

\begin{theorem}\label{distribution of statistics}
    Suppose the conditions in Theorem \ref{GA2} and Assumption \ref{sparse of R} hold. Under the null hypothesis, as $(n,p)\to\infty$, we have
    \begin{equation*}
        \mathbb{P}(T_{MAX}^{(m)}\le x)\to \exp\left\{-\frac{1}{\sqrt{\pi}}e^{-x/2}\right\}.
    \end{equation*}
\end{theorem}
According to Theorem \ref{distribution of statistics}, $H_0$ will be rejected when our proposed statistic $T_{MAX}$ is larger than the $(1-\alpha)$ quantile $q_{1-\alpha}=-\log\pi -2\log\log(1-\alpha)^{-1}$ of the Gumbel distribution $F(x)$. Next, we give the following theorem to demonstrate the consistency of our test.
\begin{theorem}\label{thpower}
    Suppose the conditions assumed in Theorem \ref{distribution of statistics} holds, for any given $\alpha\in (0,1)$, if $\|\bm{\theta}\|_\infty\ge \widetilde{C}n^{-1/2}\{\log p-2\log\log(1-\alpha)^{-1}\}^{1/2}$, for some large enough constant $\widetilde{C}$, then
    \begin{equation*}
        \mathbb{P}(T_{MAX}^{(m)}>q_{1-\alpha}|H_1)\to 1,
    \end{equation*}
    as $(n,p)\to \infty$.
\end{theorem}
With the weighted approach, our proposed test can find a better asymptotic relative efficiency. We consider a special alternative hypothesis:
\begin{equation*}
    H_1: \bm{\theta}=(\theta_1,0,\cdots,0)^\top, \theta_1>0
\end{equation*}
which means there are only one variable has nozero means. Let $x_\alpha=2\log p -\log\log p+q_{1-\alpha}$. In this case,
\begin{align*}
     &\mathbb{P}(T^{(m)}_{MAX}\ge x_\alpha)\ge  \mathbb{P}(\hat{d}^{-2}_1\hat{\theta}_1^2n\hat{\zeta}_{m-1}^2\hat{\zeta}_{2m}^{-1}p\ge x_\alpha),\\
   &\mathbb{P}(T^{(m)}_{MAX}\ge x_\alpha)\le  \mathbb{P}(\hat{d}^{-2}_1\hat{\theta}_1^2n\hat{\zeta}_{m-1}^2\hat{\zeta}_{2m}^{-1}p\ge x_\alpha)+\mathbb{P}(\max_{2\le i\le p}\hat{d}_i^{-2}\hat{\theta}_i^2n\hat{\zeta}_{m-1}^2\hat{\zeta}_{2m}^{-1}p\ge x_\alpha).
\end{align*}
Under this special alternative hypothesis, we can easily have
\begin{equation*}
    \mathbb{P}(\max_{2\le i\le p}\hat{d}_i^{-2}\hat{\theta}_i^2n\hat{\zeta}_{m-1}^2\hat{\zeta}_{2m}^{-1}p\ge x_\alpha)\to \alpha,
\end{equation*}
and
\begin{equation*}
    \mathbb{P}(\hat{d}^{-2}_1\hat{\theta}_1^2n\hat{\zeta}_{m-1}^2\hat{\zeta}_{2m}^{-1}p\ge x_\alpha)\to \Phi(-\sqrt{x_\alpha}+(np)^{1/2}d_1^{-1}\theta_1{\zeta}_{m-1}{\zeta}_{2m}^{-1/2}).
\end{equation*}
So, the power function of our proposed test $T_{MAX}^{(m)}$ satisfies
\begin{equation*}
    \beta(\bm{\theta})\in(\Phi(-\sqrt{x_\alpha}+(np)^{1/2}d_1^{-1}\theta_1{\zeta}_{m-1}{\zeta}_{2m}^{-1/2}),\Phi(-\sqrt{x_\alpha}+(np)^{1/2}d_1^{-1}\theta_1{\zeta}_{m-1}{\zeta}_{2m}^{-1/2})+\alpha).
\end{equation*}
Thus, the asymptotic relative efficiency of $T_{MAX}^{(m)}$ with respective to $T_{MAX}^{(-1)}$ could be approximated as
\begin{align*}
    ARE(T_{MAX}^{(-1)},T_{MAX}^{(m)})=\mathbb{E}(R_i^{-2})\mathbb{E}(R_i^{2m})/\mathbb{E}(R_i^{m-1})^2.
\end{align*}
By H$\ddot{\text{o}}$lder inequality, we have $ARE(T_{MAX}^{(-1)},T_{MAX}^{(m)})\ge 1$. Therefore, the inverse norm max-type test $T_{MAX}^{(-1)}$ (abbreviated as IN-MAX hereafter) has the best ARE among our proposed weighted max-type test. Its power function satisfies
\begin{equation*}
    \beta_{IN-MAX}(\bm{\theta})\in(\Phi(-\sqrt{x_\alpha}+(np)^{1/2}d_1^{-1}\theta_1{\zeta}_{-2}^{1/2}),\Phi(-\sqrt{x_\alpha}+(np)^{1/2}d_1^{-1}\theta_1{\zeta}_{-2}^{1/2})+\alpha).
\end{equation*}

Note that the test proposed by \cite{cai} corresponds to \( m = 1 \), while the test introduced by \cite{Liu23} corresponds to \( m = 0 \). So, the power function of \cite{cai}’s test (abbreviated as MAX hereafter) satisfies
\begin{equation*}
    \beta_{MAX}(\bm{\theta})\in (\Phi(-\sqrt{x_\alpha}+n^{1/2}\varsigma_1^{-1}\theta_1),\Phi(-\sqrt{x_\alpha}+n^{1/2}\varsigma_1^{-1}\theta_1)+\alpha),
\end{equation*}
where $\varsigma_i^2$ is the variance of $X_{ki},i=2,\cdots,p$. And the power function of \cite{Liu23}'s test (abbreviated as SS-MAX hereafter) satisfies
\begin{equation*}
    \beta_{SS-MAX}(\bm{\theta})\in(\Phi(-\sqrt{x_\alpha}+(np)^{1/2}d_1^{-1}\theta_1{\zeta}_{-1}),\Phi(-\sqrt{x_\alpha}+(np)^{1/2}d_1^{-1}\theta_1{\zeta}_{-1})+\alpha).
\end{equation*}
The asymptotic relative efficiencies of the inverse norm max-type test $T_{MAX}^{(-1)}$ with respect to $T_{MAX}$ and $T_{SS-MAX}$ could be approximated as follows:
\begin{align*}
    ARE_{IN-MAX,MAX}&=\mathbb{E}(R_{i}^{-2})\mathbb{E}(R_i^2)\ge 1, \\
    ARE_{IN-MAX,SS-MAX}&=\mathbb{E}(R_{i}^{-2})\mathbb{E}^{-2}(R_i^{-1})\ge 1 .
\end{align*}

In the above two inequalities, the equal sign is true if and only if
$\|R_{i}\|/ {E}(\|R_{i}\|) \cp 1$, hence in such situation these
three tests are asymptotically equivalent. Otherwise, $T_{MAX}^{(-1)}$
has larger asymptotic power than the remaining two tests. Table \ref{t1}
reports the ARE results between these three tests
under some different distributions.
Table \ref{t1} suggests that in most cases, the power advantage of $T_{MAX}^{(-1)}$ over $T_{MAX}$ and $T_{SS-MAX}$ is clear, especially in case of heavy tailed distribution.

 \begin{table}[!ht]
           \centering
           \caption{The ARE results between IN-MAX, MAX and SS-MAX under some different distributions. }
           \vspace{0.1cm}
      \renewcommand{\arraystretch}{1.1}
     \tabcolsep 4pt
         \begin{tabular}{cccccccccc}\hline \hline
  & {\tiny $t_p(0,\I_p,3)$} & {\tiny $t_p(0,\I_p,4)$} &{\tiny $t_p(0,\I_p,5)$} &{\tiny
  $t_p(0,\I_p,6)$}
  & {\tiny $N(\bm 0, \I_p)$} &{\tiny ${MN}(0.2,3,\I_p)$} & {\tiny ${MN}(0.2,10,\I_p)$} &{\tiny ${MN}(0.5,10,\I_p)$}\\
  {{\rm ARE(IN-MAX,MAX)}} & 3.00 & 2.00 & 1.67 & 1.50  & 1.00 &2.25 &16.68 &25.50\\
  {\rm ARE(IN-MAX,SS-MAX)} & 1.18 & 1.13 & 1.11 & 1.09  & 1.00 &1.09 &1.19  &1.67\\
    {{\rm ARE(SS-MAX,MAX)}} & 2.54 & 1.76 & 1.51 & 1.38  & 1.00 &2.06 &13.98 &15.27\\\hline \hline
               \end{tabular}\label{t1}\\
               \vspace{0.2cm}
            \raggedright
\small{Note: $t_p(0, \Lambda, v)$ denotes a $p$-dimensional t-distribution with degrees of freedom $v$ and scatter matrix $\Lambda$.
    ${MN}(\gamma, \sigma, \Lambda)$ refers to a mixture multivariate normal distribution with density function
    $(1-\gamma) f_p(0, \Lambda) + \gamma f_p(0, \sigma^2 \Lambda)$, where $f_p(a, b)$ is the density function of the $p$-dimensional normal distribution with mean $a$ and covariance matrix $b$.}
           \end{table}

\section{Weighted maxsum test}
\subsection{Existing Sum-type test}
\cite{FLM21} proposed the following weighted sum-type test statistic:
\begin{equation*}
    T_{SUM}(w)=\frac{2}{n(n-1)}{\sum\sum}_{i<j}w\left(\|\widetilde{\bm{D}}^{-1/2}_{ij}\bm{X}_i\|\right)w\left(\|\widetilde{\bm{D}}^{-1/2}_{ij}\bm{X}_j\|\right)U\left(\widetilde{\bm{D}}^{-1/2}_{ij}\bm{X}_i\right)^\top U\left(\widetilde{\bm{D}}^{-1/2}_{ij}\bm{X}_i\right),
\end{equation*}
where $\widetilde{D}_{i,j}$ are the corresponding diagonal matrix estimator in (\ref{oldm}) using leave-two-out sample $\{X_k\}_{k\neq i}^n$.

In order to fit our assumptions, we rewrite the above statistic as
\begin{equation*}
   T_{SUM}^{(m)}=\frac{2}{n(n-1)}{\sum\sum}_{i<j}\|\widetilde{\bm{D}}^{-1/2}_{ij}\bm{X}_i\|^m\|\widetilde{\bm{D}}^{-1/2}_{ij}\bm{X}_j\|^mU\left(\widetilde{\bm{D}}^{-1/2}_{ij}\bm{X}_i\right)^\top U\left(\widetilde{\bm{D}}^{-1/2}_{ij}\bm{X}_i\right).
\end{equation*}
\begin{assumption}\label{elliptical}
    Let $X_1,\ldots,X_n$ be a sequence of $p$-dimensional independent and identically distributed (iid) observations with an elliptically symmetric density $\det(\boldsymbol{\Sigma})^{-1/2}g\{\|\boldsymbol{\Sigma}^{-1/2}(\boldsymbol{x}-\boldsymbol{\bm{\mu}})\|\}$, where the symmetry center $\boldsymbol{\bm{\mu}}\in\mathbb{R}^p$ is the location parameter, the symmetric positive-definite matrix $\boldsymbol{\Sigma}\in\mathbb{R}^{p\times p}$ is the scale parameter, and $g$ is a univariate probability density function. Define the diagonal elements of $\mathbf{\Sigma}$ are $\lambda_1, \lambda_2,\dots,\lambda_p$. $0<\underline{\lambda}<\lambda_i<\bar{\lambda}$, $i=1,2,\dots,p$ for some constants $\underline{\lambda}$ and $\bar{\lambda}$
\end{assumption}
\begin{assumption}\label{R1}
    (i) $\mathrm{tr}(\mathbf{R}^4)=o(\mathrm{tr}^2(\mathbf{R}^2))$; (ii) $\mathrm{tr}(\mathbf{R}^2)-p=o(n^{-1}p^2)$ and $\log p =o(n)$
\end{assumption}
\begin{assumption}
    $\zeta_{4m}=O(\zeta_{2m}^2)$; for $k=1,2,3,4$ exist for large enough $p$ and for $k=2,3,4$, $\zeta_{-k}/\zeta^k_{-1}\to \gamma_k\in \left[1,+\infty\right)$ as $p\to \infty$, where $\gamma_k$'s are constants.
\end{assumption}
The following Lemma restates Theorem 1 in \cite{FLM21}, which provides the asymptotic null distribution of $T_{SUM}^{(m)}$ under the assumption of a symmetric elliptical distribution.
\begin{lemma}
    Under Assumption 8-10, and $H_0$, as $\min(p,n)\to \infty$, $$T_{SUM}^{(m)}/\sigma_n\overset{d}{\to}N(0,1),$$ where $\sigma_n^2=2n^{-2}p^{-2}\zeta_{2m}^2\mathrm{tr}(\mathbf{R}^2)$,
\end{lemma}

To expand the scope of application, we re-derive the limiting null distribution of $T_{SUM}^{(m)}$ under a more general model  (\ref{model}).
\begin{theorem}\label{sizesum}
    Under Assumption 2-5, 9-10 and $H_0$,as $(n,p)\to \infty$, $$T_{SUM}^{(m)}/\sigma_n\overset{d}{\to}N(0,1),$$ where $\sigma_n^2=2n^{-2}p^{-2}\zeta_{2m}^2\mathrm{tr}(\mathbf{R}^2)$.
\end{theorem}
In practice, $\sigma^2_n$ can be consistently estimated as
$$\begin{aligned}
\widehat{\sigma_{n}^{2}}=&2n^{-4}{\sum\sum}_{i\neq j}\|\widetilde{\mathbf{D}}_{ij}^{-1/2}\mathbf{X}_{i}\|^{2m}\|\widetilde{\mathbf{D}}_{ij}^{-1/2}\mathbf{X}_{j}\|^{2m}\left\{U(\widetilde{\mathbf{D}}_{ij}^{-1/2}\mathbf{X}_{i})-\tilde{\boldsymbol{\mu}}_{i,j}\right\}^{\mathrm{T}}U(\widetilde{\mathbf{D}}_{ij}^{-1/2}\mathbf{X}_{j})\\
&\times\left\{U(\widetilde{\mathbf{D}}_{ij}^{-1/2}\mathbf{X}_{j})-\tilde{\boldsymbol{\mu}}_{i,j}\right\}^{\mathrm{T}}U(\widetilde{\mathbf{D}}_{ij}^{-1/2}\mathbf{X}_{i}).
\end{aligned}$$
Thus, we reject the null hypothesis when $T_{SUM}^{(m)}/\sqrt{\widehat{\sigma_n^2}}>$ $z_\alpha$, where $z_\alpha$ is the upper $\alpha$-quantile of the standard normal distribution $N(0,1)$ for a level $\alpha$ test.

Next, we consider the following alternative hypothesis:
\begin{equation}\label{ha}
H_1:\bm{\theta}^\top\mathbf{D}^{-1}\bm{\theta}=O(n^{-1}\sqrt{\mathrm{tr}(\mathbf{R}^2)})\;\text{~and~}\; \bm{\theta}^\top\mathbf{D}^{-1/2}\mathbf{R}\mathbf{D}^{-1/2}\bm{\theta}=o(n^{-1}\mathrm{tr}(\mathbf{R}^2)).
\end{equation}
\begin{theorem}\label{powersum}
     Under Assumptions 2-5, 9-10 and the alternative hypothesis (\ref{ha}),as $\min(p,n)\to \infty$, $$\frac{T_{SUM}^{(m)}-\zeta_{m-1}^2\bm{\theta}\mathbf{D}^{-1}\bm{\theta}}{\sigma_n}\overset{d}{\to}N(0,1).$$
\end{theorem}
By Theorem \ref{sizesum} and Theorem \ref{powersum}, we have the asymptotic power function of $T_{SUM}^{(m)}$ is
\begin{equation*}
    \beta(\bm{\theta})=\Phi\left(-z_\alpha+\frac{\zeta_{m-1}^2pn\bm{\theta}^\top\mathbf{D}^{-1}\bm{\theta}}{\zeta_{2m}\sqrt{2\mathrm{tr}(\mathbf{R}^2)}}\right).
\end{equation*}
It is easy to find the inverse norm sign test $T_{SUM}^{(-1)}$ is more powerful than $T_{SUM}^{(m)}$ for $m\neq -1$.
\subsection{Weighted maxsum test}
In this subsection, we demonstrate that the proposed weighted max-type statistic $T_{MAX}^{(m)}$ is asymptotically independent of the weighted sum-type statistic $T_{SUM}^{(m)}$. This property enables us to perform a Cauchy $p$-value combination of the two asymptotically independent $p$-values, thereby constructing a new test. This newly developed test is designed to handle both sparse and dense alternatives effectively.
\begin{assumption}
    There exist $C>0$ and $\varrho\in(0,1)$ so that $\max_{1\leq i<j\leq p}|\sigma_{ij}|\leq\varrho$ and $\max_{1\leq i\leq p}\sum_{j=1}^p\sigma_{ij}^2\leq( \log p) ^{C}$ for all $p\geq 3$ ; $ p^{- 1/ 2}( \log p) ^{C}\ll \lambda _{\min }( \mathbf{R} ) \leq \lambda _{\max }( \mathbf{R} ) \ll \sqrt {p}( \log p) ^{- 1}$ and $\lambda _{\max }( \mathbf{R} ) / \lambda _{\min }( \mathbf{R} ) =O\left ( p^{\tau }\right )$ for some $\tau \in ( 0, 1/ 4) .$
\end{assumption}
\begin{remark}
   Assumption 11 is identical to condition (2.3) in \cite{F22a}. As indicated in \cite{F22a}, Assumption 11 is more restrictive than Assumption 5, Assumption 7, and part (i) of Assumption 9. Under Assumption 11, we have $p^{1/2}(\log p)^C\lesssim \text{tr}(\mathbf{R}^2)\lesssim p^{3/2}\log^{-1}p$. Consequently, Assumption 6 will be satisfied if $n = o(p^{3/2}\log p)$ and $p^{3/4}(\log p)^{-C/2}=O(n)$. Intuitively, when all the eigenvalues of $\mathbf{R}$ are bounded and $\frac{p}{n}\to c\in(0,\infty)$, all of Assumptions 5, 6, 7, and 9 hold.
\end{remark}
\begin{theorem}\label{a.i.1}
    Under Assumptions 1-7, 9-10 and $H_0$, $T_{SUM}^{(m)}$ and $T_{MAX}^{(m)}$ are asymptotically independent, i.e.
    \begin{equation*}
        \mathbb{P}\left(T_{SUM}^{(m)}/\sigma_n\le x, T_{MAX}^{(m)}\le y\right)\to \Phi(x)F(y),
    \end{equation*}
    as $(n,p)\to \infty$
\end{theorem}
Based on Theorem \ref{a.i.1}, we propose combining the corresponding $p$-values by means of the Cauchy Combination Method \citep{LX}, that is,
$$\begin{aligned}
&p_{CC}=1-G\left[0.5\tan\left\{(0.5-p^{(m)}_{MAX})\pi\right\}+0.5\tan\left\{(0.5-p^{(m)}_{SUM})\pi\right\}\right],\\
&p^{(m)}_{MAX}=1-F(T^{(m)}_{MAX}),~~ p^{(m)}_{SUM}=1-\Phi\left(\frac{T^{(m)}_{SUM}}{\hat{\sigma}_{n}}\right),
\end{aligned}$$
where $G(\cdot)$ represents the cumulative distribution function (CDF) of the standard Cauchy distribution. If the final $p$-value is smaller than a pre-specified significance level $\alpha\in(0,1)$, then we reject the null hypothesis $H_0$.

Subsequently, we examine the relationship between $T^{(m)}_{SUM}$ and $T^{(m)}_{MAX}$ under local alternative hypotheses:
\begin{equation}\label{H1cc}
    H_{1}:\|\boldsymbol{\theta}\| = O\left(n^{-1}\sqrt{\mathrm{tr}(\mathbf{R}^2)}\right),\|\mathbf{R}^{1/2}\boldsymbol{\theta}\| = o\left(n^{-1}{\mathrm{tr}(\mathbf{R}^2)}\right),|\mathcal{A}| = o\left(\frac{\lambda_{\min}(\mathbf{R})[\mathrm{tr}(\mathbf{R}^{2})]^{1/2}}{(\log p)^{C}}\right),
\end{equation}
where $\mathcal{A}=\{i\mid\theta_i\neq0,1\leq i\leq p\}$ and $\boldsymbol{\theta}=(\theta_1,\theta_2,\cdots,\theta_p)^\top$. The following theorem establishes the asymptotic independence between $T^{(m)}_{SUM}$ and $T^{(m)}_{MAX}$ under this particular alternative hypothesis.
\begin{theorem}\label{a.i.2}
    Under Assumptions 1-7, 9-10 and the alternative hypothesis \ref{ha}, $T_{SUM}^{(m)}$ and $T_{MAX}^{(m)}$ are asymptotically independent i.e.
    \begin{equation*}
        \mathbb{P}\left(T_{SUM}^{(m)}/\sigma_n\le x, T_{MAX}^{(m)}\le y\right)\to\mathbb{P}\left(T_{SUM}^{(m)}/\sigma_n\le x\right)\mathbb{P}\left(T_{MAX}^{(m)}\le y\right)
    \end{equation*}
    as $(n,p)\to \infty$.
\end{theorem}
Based on the findings in \cite{L23}, the test founded on the Cauchy combination exhibits greater power than the test relying on the minimum of $p^{(m)}_{MAX}$ and $p^{(m)}_{SUM}$, also recognized as the minimal p-value combination. This is expressed as $\beta^{(m)}_{M\wedge S,\alpha}=\mathbb{P}(\min \{p^{(m)}_{MAX},p^{(m)}_{SUM}\}\leq1-\sqrt{1-\alpha})$.

It is evident that:
\begin{equation}\label{b1}
    \begin{aligned}
    \beta^{(m)}_{M\Lambda S,\alpha}&\geq P(\min\{{p}^{(m)}_{MAX},{p^{(m)}}_{SUM}\}\leq\alpha/2)\\
    &=\beta^{(m)}_{MAX,\alpha/2}+\beta^{(m)}_{SUM,\alpha/2}-P({p}^{(m)}_{MAX}\leq\alpha/2,{p}^{(m)}_{SUM}\leq\alpha/2)\\
    &\geq\max\{\beta^{(m)}_{MAX,\alpha/2},\beta^{(m)}_{SUM,\alpha/2}\}.
    \end{aligned}
\end{equation}

Conversely, under the local alternative hypothesis (\ref{H1cc}), we obtain:
\begin{equation}\label{b2}
    \beta^{(m)}_{M\wedge S,\alpha}\geq\beta^{(m)}_{MAX,\alpha/2}+\beta^{(m)}_{SUM,\alpha/2}-\beta^{(m)}_{MAX,\alpha/2}\beta^{(m)}_{SUM,\alpha/2},
\end{equation}
which can be attributed to the asymptotic independence indicated by Theorem \ref{a.i.2}.

For a small $\alpha$, the disparity between $\beta^{(m)}_{MAX,\alpha}$ and $\beta^{(m)}_{MAX,\alpha/2}$ is negligible, and the same holds for $\beta^{(m)}_{SUM,\alpha}$. Thus, in light of equations (\ref{b1}) and (\ref{b2}), the power of the adaptive test is at least as great as, or even substantially greater than, that of either the max-type or sum-type test. For a comprehensive comparison of the performance of each test type under different conditions of sparsity and signal strength, consult Table 1 in \cite{M24}.
\section{Simulation}
In this section, we incorporated various methods into our study:
\begin{itemize}
    \item max-type methods proposed by \cite{cai} referred as MAX;
    \item sum-type methods proposed by \cite{Sri} referred as SUM;
    \item combination test proposed by \cite{F22a} referred as CC;
    \item max-type methods proposed by \cite{L24} referred as SS-MAX;
    \item sum-type methods proposed by \cite{FS} referred as SS-SUM;
    \item combination test proposed by \cite{L24} referred as SS-CC;
    \item the proposed test $T_{MAX}^{(-1)}$ referred as IN-MAX;
    \item sum-type weighted methods proposed by \cite{FLM21} referred as IN-SUM;
    \item the proposed test $T_{CC}^{(-1)}$ referred as IN-CC.
\end{itemize}
We consider the following four commonly studied simulation settings:
\begin{enumerate}[i.]
    \item Multivariate normal distribution: $N(\bm{\mu},\bm{\Sigma})$, $\bm{\Sigma}=\left(0.5^{|i-j|}\right)_{1\le i,j\le p}$;
    \item Multivariate t-distribution: $t_4(\bm{\mu},\bm{\Sigma})$, $\bm{\Sigma}=\left(0.5^{|i-j|}\right)_{1\le i,j\le p}$;
    \item Mixed multivariate normal distribution: $MN(0.8,3,\bm{\mu},\bm{\Sigma})$, $\bm{\Sigma}=\left(0.5^{|i-j|}\right)_{1\le i,j\le p}$
    \item Mixed multivariate normal distribution: $MN(0.2,3,\bm{\mu},\bm{\Sigma})$, $\bm{\Sigma}=\left(0.5^{|i-j|}\right)_{1\le i,j\le p}$.
\end{enumerate}
We set the sample size at $n = 80$, while the dimension takes on values of $p = 200, 400, 600$ respectively. All the results in this section are obtained from 1000 repetitions. Table 1 showcases the empirical sizes of the nine tests previously mentioned. It was noted that the spatial-sign based tests can effectively control the empirical sizes in most scenarios. However, the SUM test is too conservative under non-normal distributions.
\begin{table}[ht]
\centering
\caption{Empirical sizes of MAX, SUM, CC, SS-MAX, SS-SUM, SS-CC, IN-MAX, IN-SUM and IN-CC at 5\% level under Settings i--iv.}
\begin{tabular}{cccccccccc}
\toprule
Setting & MAX & SUM & CC & SS-MAX & SS-SUM & SS-CC & IN-MAX & IN-SUM & IN-CC \\
\midrule
\multicolumn{9}{c}{$(n, p) = (80, 200)$} \\
\midrule
i & 0.069 & 0.066 & 0.070 & 0.038 & 0.067 & 0.069 & 0.044 & 0.068 & 0.077 \\
ii & 0.048 & 0.003 & 0.026 & 0.038 & 0.075 & 0.067 & 0.046 & 0.079 & 0.083 \\
iii & 0.044 & 0.006 & 0.019 & 0.037 & 0.066 & 0.063 & 0.042 & 0.075 & 0.067 \\
iv & 0.068 & 0.055 & 0.054 & 0.026 & 0.053 & 0.048 & 0.055 & 0.064 & 0.073 \\
\midrule
\multicolumn{9}{c}{$(n, p) = (80, 400)$} \\
\midrule
i & 0.083 & 0.057 & 0.071 & 0.048 & 0.060 & 0.058 & 0.049 & 0.058 & 0.063 \\
ii & 0.036 & 0.000 & 0.012 & 0.039 & 0.056 & 0.059 & 0.052 & 0.058 & 0.064 \\
iii & 0.044 & 0.000 & 0.026 & 0.033 & 0.054 & 0.054 & 0.041 & 0.055 & 0.053 \\
iv & 0.057 & 0.030 & 0.038 & 0.023 & 0.050 & 0.041 & 0.060 & 0.068 & 0.071 \\
\midrule
\multicolumn{9}{c}{$(n, p) = (80, 600)$} \\
\midrule
i & 0.091 & 0.062 & 0.081 & 0.047 & 0.061 & 0.064 & 0.048 & 0.058 & 0.063 \\
ii & 0.047 & 0.000 & 0.029 & 0.046 & 0.064 & 0.063 & 0.050 & 0.066 & 0.070 \\
iii & 0.025 & 0.000 & 0.012 & 0.038 & 0.057 & 0.053 & 0.043 & 0.057 & 0.055 \\
iv & 0.077 & 0.031 & 0.040 & 0.031 & 0.057 & 0.050 & 0.048 & 0.056 & 0.055 \\
\bottomrule
\end{tabular}
\end{table}

\begin{figure}[htbp]
    \centering
    \includegraphics[width=0.8\linewidth]{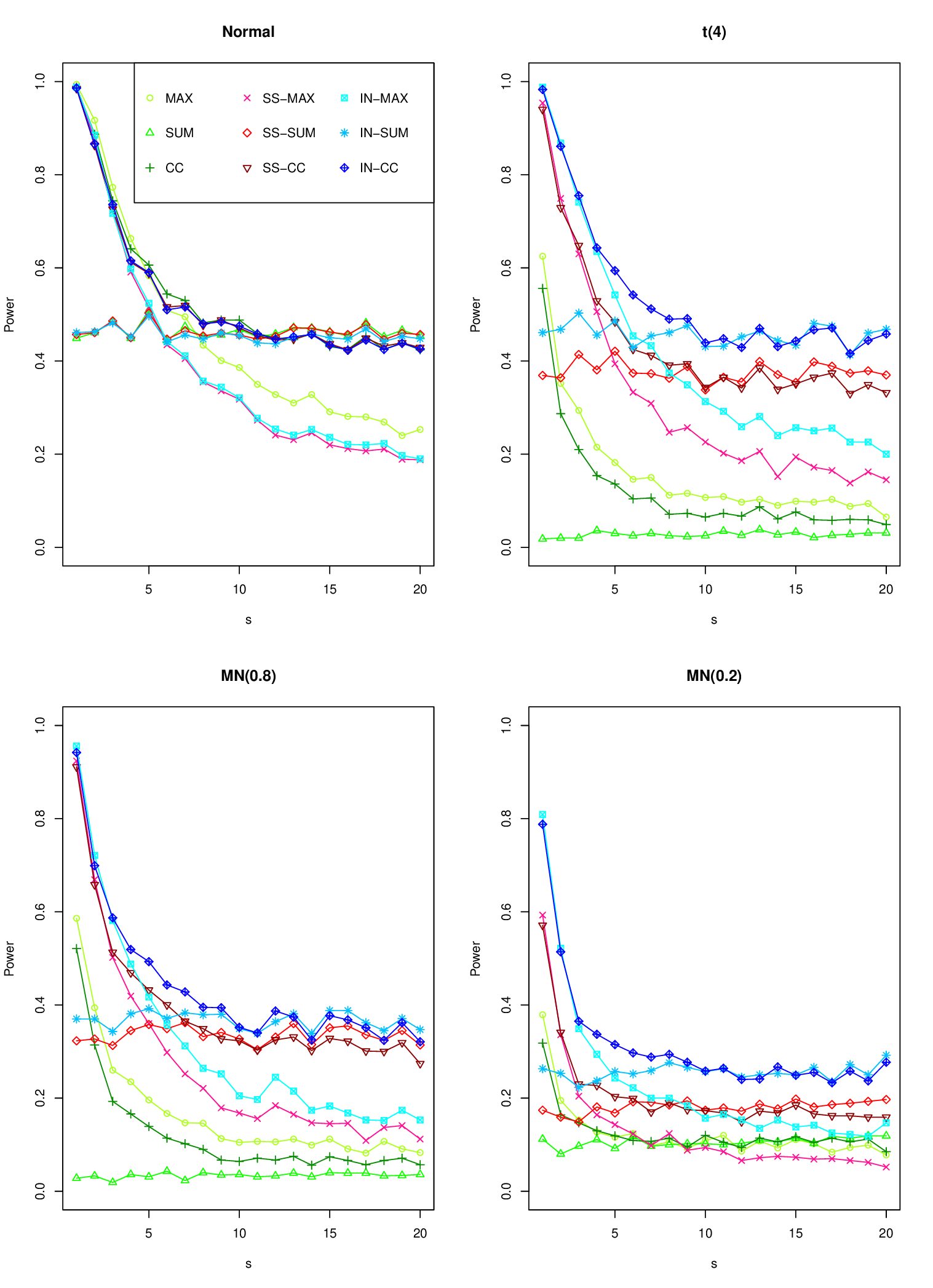}
    \caption{Power of tests with different sparsity levels over $(n,p)=(80,200)$}
 \label{p-s}
\end{figure}

To demonstrate the comparison among the various tests, we consider $\bm{\theta}=(\kappa,\cdots,\kappa,0,\cdots,0)$, where the first $s$ components of $\bm{\theta}$ are equal to $\kappa = \sqrt{\delta/s}$. Figure \ref{p-s} depicts the power of the tests with different sparsity levels for $(n,p)=(80,200)$ under different distributions. In Settings i, ii and iii, we present the plots for $\delta = 0.5$, while in Setting iv, we show the plots for $\delta = 1$.

We observe that IN-MAX, IN-SUM, and IN-CC are the most powerful within the max-type tests, sum-type tests, and combination tests, respectively, especially in the context of non-normal distributions. Moreover, the newly proposed test, IN-CC, showcases superior performance compared to other methods across most scenarios. In scenarios characterized by extremely sparse signals ($s < 5$), the performance of IN-CC is on par with that of IN-MAX. In highly dense scenarios ($s > 10$), IN-CC displays performance similar to IN-SUM. However, in situations where the signal density is neither extremely low nor extremely high, IN-CC proves to be the most effective test procedure among all those evaluated. This underscores the advantages of our proposed max-sum procedures, particularly in handling signal sparsity and dealing with heavy-tailed distributions.

Subsequently, we examine the power comparison of these tests under different signal strengths. Here, we consider three sparsity levels: $s = 2, 10, 20$, and the signal parameter $\kappa=\sqrt{\delta/s}$. Figures 2 to 4 display the power curves for various testing methods in Settings i-iv. As the signal strength rises, the power of all tests increases accordingly. Even in the presence of heavy-tailed distributions, the spatial-sign based testing methods continue to outperform those based on means. Among all the tests, the proposed IN-CC test consistently exhibits the best performance.
\begin{figure}[!ht]
	\centering
    \includegraphics[width=\linewidth]{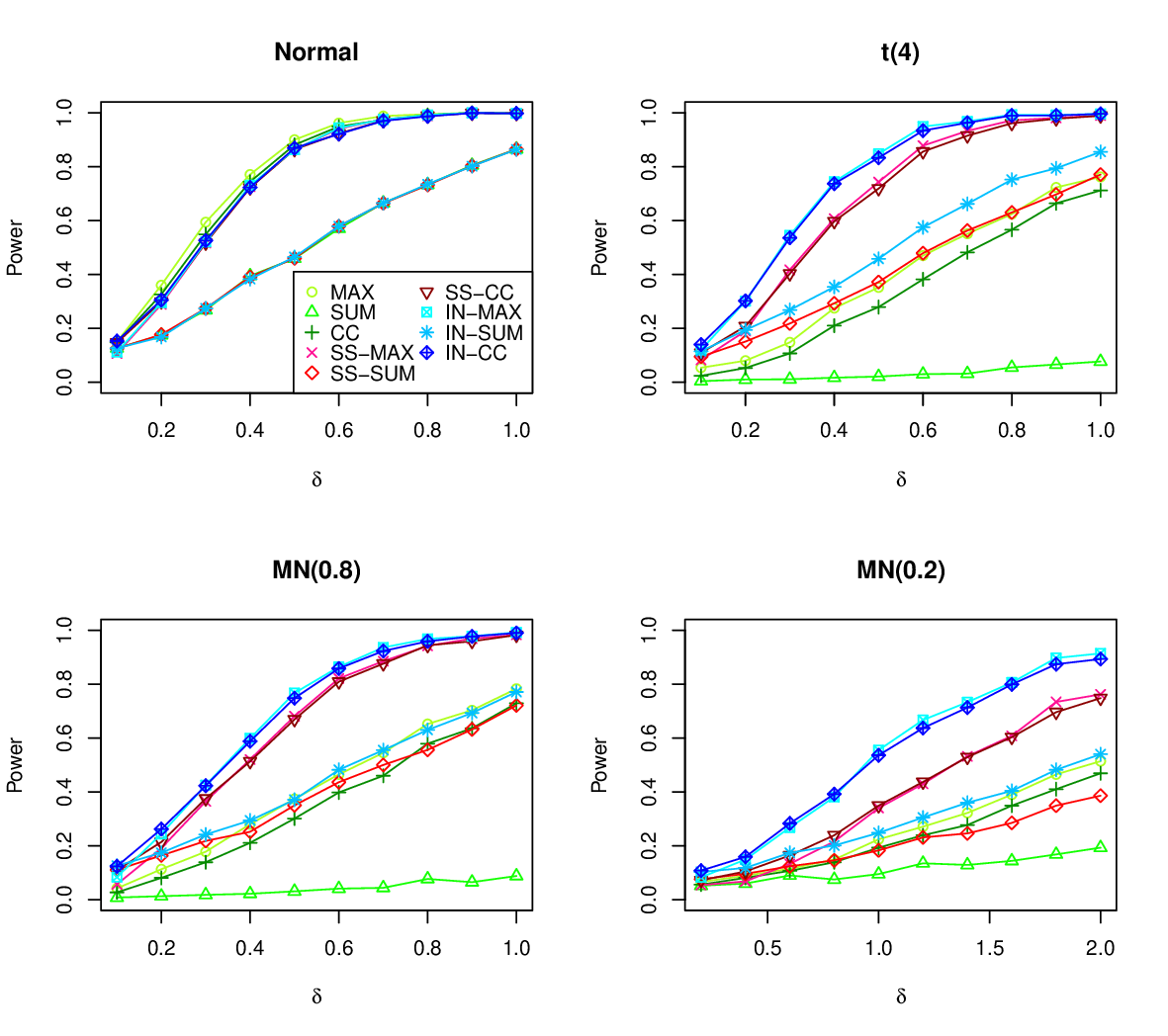}
 \caption{Power of tests with different distribution for $s=2$ over $(n,p)=(80,200)$}
 \label{p-d1}
\end{figure}
\begin{figure}[!ht]
	\centering
	\includegraphics[width=\linewidth]{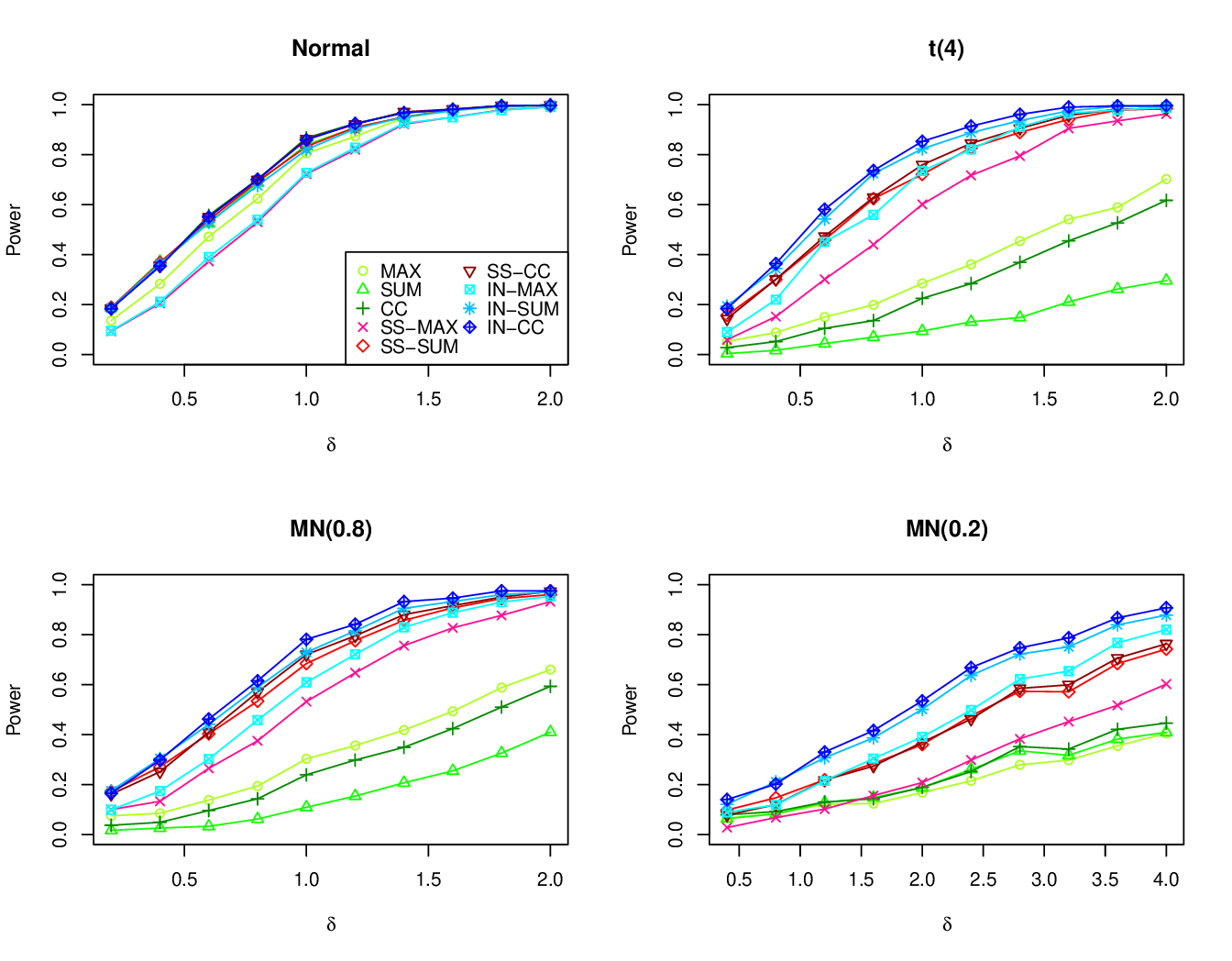}
 \caption{Power of tests with different distribution for $s=10$ over $(n,p)=(80,200)$}
 \label{p-d2}
\end{figure}
\begin{figure}[!ht]
	\centering
		\includegraphics[width=\linewidth]{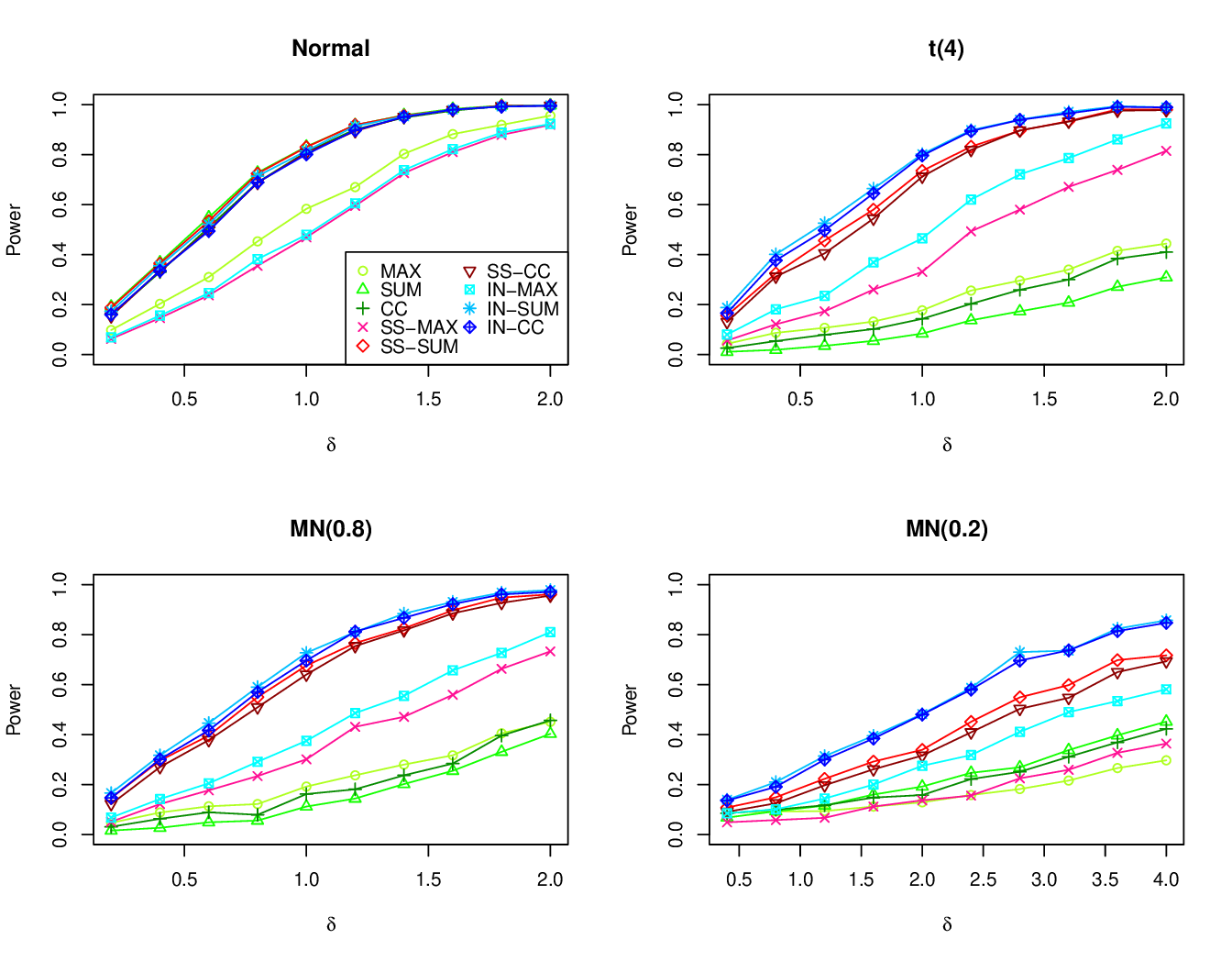}
 \caption{Power of tests with different distribution for $s=20$ over $(n,p)=(80,200)$}
 \label{p-d3}
\end{figure}
\section{Application}
In this section, we utilize our methods to tackle a financial pricing problem. The aim is to determine whether the expected returns of all assets are equivalent to their respective risk-free returns. Let $X_{ij}\equiv R_{ij}-\text{rf}_i$ represent the excess return of the $j$-th asset at time $i$, where $i = 1,\cdots,n$ and $j = 1,\cdots,p$. Here, $R_{ij}$ stands for the return on asset $j$ in period $i$, and $\text{rf}_i$ is the risk-free return rate of all assets during period $i$. Subsequently, we investigate the following pricing model:
\begin{equation}\label{model5}
    X_{ij}=\mu_j+\xi_{ij},
\end{equation}
for $i = 1,\cdots,n$ and $j = 1,\cdots,p$. In vector form, this can be expressed as $\mathbf{X}_i=\boldsymbol{\mu}+\boldsymbol{\xi}_i$, where $\mathbf{X}_i=(X_{i1},\ldots,X_{ip})^\top$, $\boldsymbol{\mu}=(\mu_1,\ldots,\mu_p)^\top$, and $\boldsymbol{\xi}_i = (\xi_{i1},\ldots,\xi_{ip})^\top$ is the zero-mean error vector. We consider the following hypothesis:
\begin{equation*}
    H_0:\boldsymbol{\mu}=\boldsymbol{0}\quad\text{versus}\quad H_1:\boldsymbol{\mu}\neq\boldsymbol{0}.
\end{equation*}

We analyzed the weekly return rates of stocks within the S\&P 500 index from January 2005 to November 2018. The weekly data were sourced from the stock prices recorded every Friday. During this period, the composition of the index changed, with some stocks being added. As a result, we concentrated on a total of 424 stocks that were continuously part of the S\&P 500 index throughout. For each of these stocks, we collected 716 weekly return rates, excluding Fridays that were holidays.

Given the likelihood of autocorrelation in weekly stock returns, we applied the Ljung-Box test \cite{LB} at a 0.05 significance level to test for zero autocorrelations in each stock. We retained 280 stocks for which the Ljung-Box test at a 0.05 level was not rejected. It is important to note that if all 424 stocks had been used, there might be autocorrelation between observations. This would violate our assumption and potentially necessitate further research.

\begin{figure}[!ht]
    \centering
    \includegraphics[width=0.7\linewidth]{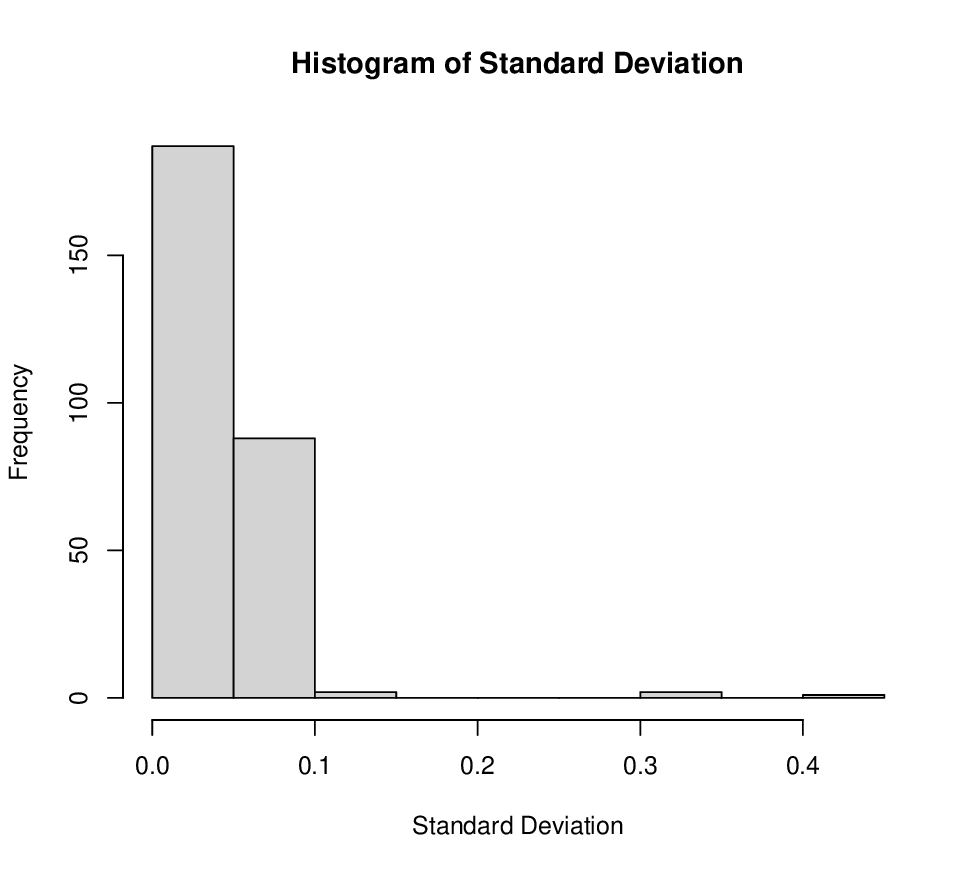}
    \caption{Histogram of standard deviation of US securities.}
    \label{std}
\end{figure}

\begin{figure}[!ht]
    \centering
    \includegraphics[width=\linewidth]{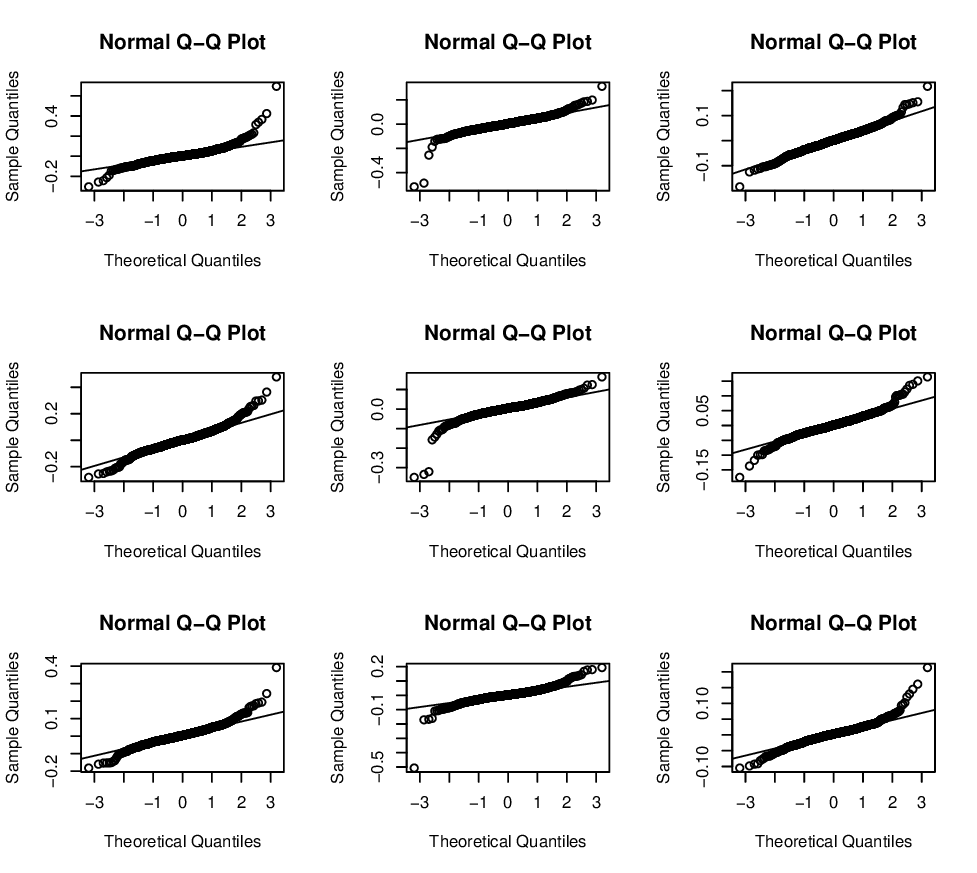}
    \caption{Q-Q plots of the weekly return rates of some stocks with heavy-tailed distributions in the S\&P 500 index.}
    \label{qq}
\end{figure}

\begin{figure}[!ht]
    \centering
    \includegraphics[width=0.7\linewidth]{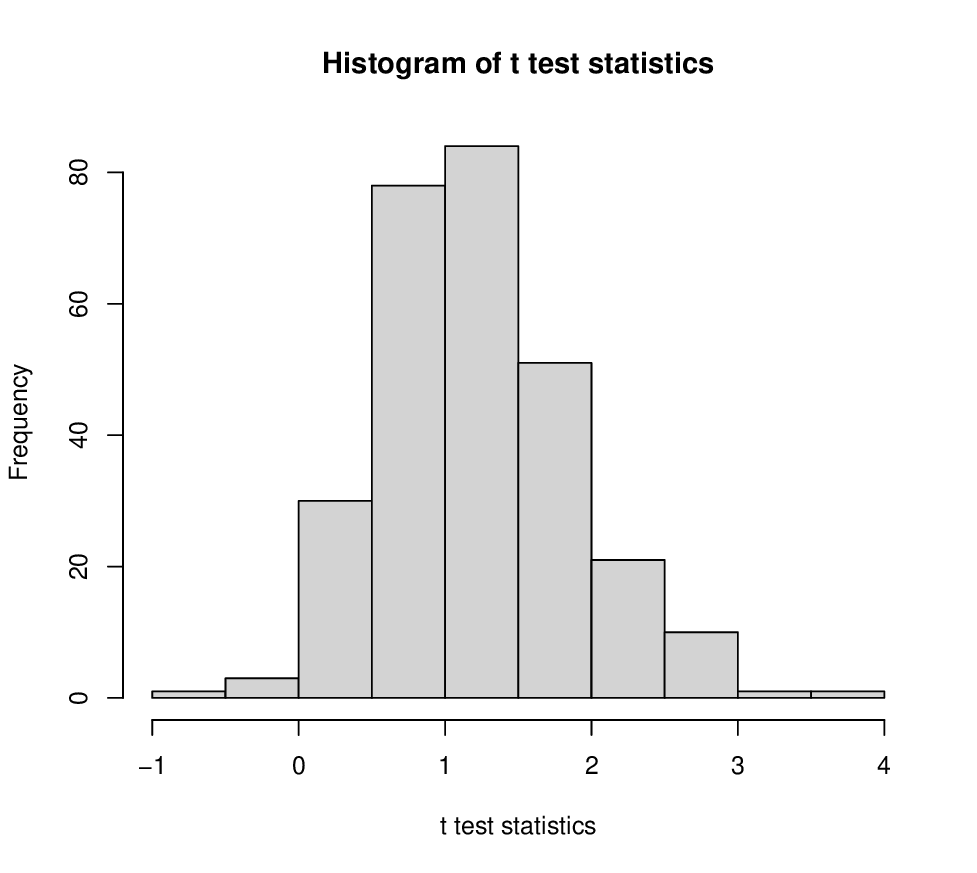}
    \caption{t test statistics of the weekly excess return rates of each stock.}
    \label{ttest}
\end{figure}

Figure \ref{std} depicts the histogram of the standard deviation of the 280 securities. It was observed that the variances of these assets are evidently unequal. Consequently, the scalar-invariant test procedure is favored. As a result, the nine aforementioned test procedures (MAX, SUM, CC, SS-SUM, SS-MAX, SS-CC, IN-MAX, IN-SUM, IN-CC) were implemented on the complete sample. All the tests convincingly reject the null hypothesis. To evaluate the effectiveness of the proposed test in comparison with other competing tests for different sample sizes, $n = 52K$ observations were randomly drawn from 716 weekly returns, where $K$ ranged from 3 to 8. The experiment was repeated 1000 times for each value of $n$. Table \ref{tab2} shows the rejection rates of the nine tests. Figure \ref{qq} presents the Q-Q plots of the weekly return rates of some stocks in the S\&P 500 index. It was noted that all the data deviate from a normal distribution and display heavy tails. Additionally, sum-type test procedures perform better than max-type test procedures, primarily due to the density of the alternative. Figure \ref{ttest} illustrates the t-test statistic for each stock. It was found that many t-test statistics are greater than 2, and the majority of them are positive. Among these tests, it can be seen that inverse norm weighted test procedures are more powerful than spatial-sign based test procedures and mean-based test procedures. Moreover, the Cauchy Combination is more powerful than both max-type and sum-type tests. The proposed IN-CC test proves to be the most powerful. Therefore, the application of real-world data also demonstrates the superiority of the proposed maxsum test procedure.
\begin{table}[!ht]
\caption{The rejection rates of testing excess returns of the S\&P stocks for $p = 280$ and $n = 52K$ with $K = 3,\cdots, 8.$ For each $n$, we sampled $1000$ data sets.}
\centering
\begin{tabular}{ ccccccc }
\hline
  & $n=156$ & $n=208$ & $n=260$ & $n=312$ & $n=364$ & $n=416$\\
\hline
MAX    & 0.124 & 0.129 & 0.141 & 0.150 & 0.165 & 0.162\\
SUM    & 0.192 & 0.239 & 0.277 & 0.264 & 0.325 & 0.343\\
CC     & 0.177 & 0.227 & 0.240 & 0.238 & 0.295 & 0.300\\
SS-MAX & 0.231 & 0.325 & 0.387 & 0.434 & 0.497 & 0.562\\
SS-SUM & 0.385 & 0.503 & 0.606 & 0.682 & 0.755 & 0.816\\
SS-CC  & 0.386 & 0.489 & 0.595 & 0.647 & 0.739 & 0.797\\
IN-MAX & 0.365 & 0.476 & 0.547 & 0.611 & 0.686 & 0.746\\
IN-SUM & 0.424 & 0.545 & 0.657 & 0.739 & 0.813 & 0.870\\
IN-CC  & 0.437 & 0.558 & 0.660 & 0.743 & 0.814 & 0.873\\
\hline
\end{tabular}
\label{tab2}
\end{table}



\section{Conclusion}
For the high-dimensional one-sample location testing problem under heavy-tailed distributions, we first propose a max-type test procedure based on the weighted spatial-sign, which exhibits strong performance, particularly in scenarios with sparse alternatives. Furthermore, we show that the optimal weighted function for this max-type test is identical to the one used in the sum-type test procedures described in \citep{FLM21}. Building on this, we establish the asymptotic independence between the proposed max-type test statistic and the sum-type test statistic outlined in \cite{FLM21}. Leveraging this asymptotic independence, we construct a combined max-sum test procedure using the Cauchy combination method. The resulting IN-CC test demonstrates excellent performance across a wide range of distributions and varying levels of sparsity in the alternatives. Notably, it is of great interest to extend this approach to other high-dimensional testing problems, such as the two-sample location problem \citep{H23} and the alpha test in linear pricing models \citep{Liu23}, opening up avenues for further research and application.

\section{Appendix}
Recall that $\mathbf{D}=$diag$\{ d_{1}^{2}, d_{2}^{2}, \cdots , d_{p}^{2}\} .$ For $i=1,2,\cdots,n,\boldsymbol{U}_{i}=U(\mathbf{D}^{-1/2}(\boldsymbol{X}_{i}-\boldsymbol{\theta}))$ and $R_{i}$ = $\| \mathbf{D} ^{- 1/ 2}( \boldsymbol{X}_{i}$ $\boldsymbol{\theta }) \|$ as the scale-invariant spatial-sign and radius of $\boldsymbol{X}_{i}-\boldsymbol{\theta}$, where $U(\boldsymbol{X})=\boldsymbol{X}/\|\boldsymbol{X}\|\mathbb{I}(\boldsymbol{X}\neq0)$ is the multivariate sign function of $\boldsymbol{X}$, with $\mathbb{I}(\cdot)$ being the indicator function. The moments of $R_i$ is defined as $\zeta_k=\mathbb{E}\left(R_{i}^{-k}\right)$.

We denote the estimated version $\boldsymbol{U}_i$ and $R_i$ as $\hat{R}_i=\|\hat{\mathbf{D}}^{-1/2}(\boldsymbol{X}_i\boldsymbol{-\theta})\|$ and $\hat{U}_i=$
$\|\hat{\mathbf{D}}^{-1/2}(\boldsymbol{X}_i\boldsymbol{-}\boldsymbol{\theta})/\|\hat{\mathbf{D}}^{\boldsymbol{-}1/2}(\boldsymbol{X}_i\boldsymbol{-}\boldsymbol{\theta})\|$, respectively$,i=1,2,\cdots,n.$
\subsection{The lemmas to be used}
\begin{lemma}\label{order of varU}
    Under Assumption \ref{struction of W}, we have $\mathbb{E}(U(\bm{W}_i)^\top\bm{M}U(\bm{W}_i)^2)=O(p^{-2}\mathrm{tr}(\bm{M}^\top\bm{M}))$.
\end{lemma}
\begin{lemma}\label{Convergence of D}
    Under Assumptions \ref{existence and uniqueness of HR}, \ref{struction of W}, \ref{weighted function}, \ref{R2}, if $\zeta_{2m}=O(\zeta_m^2)$ we have $\max_{j=1,\dots,p}(\hat{d}_j-d_j)=O_p(n^{-1/2}(\log p)^{1/2})$.
\end{lemma}
\begin{lemma}\label{Qh}
    Under Assumptions \ref{existence and uniqueness of HR}, \ref{struction of W}, \ref{weighted function}, \ref{R2} and $\zeta_{2m}=O(\zeta_m^2)$. Define a random $p\times p$ matrix $\hat{\mathbf{Q}}(s)=n^{-1}\sum_{i=1}^n \hat{R}_i^s\hat{\bm{U}}_i\hat{\bm{U}}^\top_i$ and let $\hat{\mathbf{Q}}_{jl}$ be the $(j,l)$th element of $\hat{\mathbf{Q}}$. If there exist two positive constants $\underline{b}$ and $\bar{B}$ such that $\underline{b}\le \lim\sup_p \mathbb{E}(R_i/\sqrt{p})^{k}\le \bar{B}$ for $k\in \{s,2s,3s,4s,s-2,2s-4\}$, then
    \begin{equation*}
     |\hat{\mathbf{Q}}_{jl}|\lesssim \zeta_s p^{-1}|\sigma_{jl}|+O_p\left( \zeta_s n^{-1/2}p^{-1}+\zeta_s p^{-7/6}+\zeta_s p^{-1-\delta/2}+\zeta_s n^{-1/2}(\log p)^{1/2}(p^{-5/2}+p^{-1-\delta/2} )\right).
\end{equation*}
\end{lemma}
\begin{lemma}\label{777}
     Under Assumptions 1-6, and $a_0(p)\asymp p^{1-\delta}$ for some positive constant $\delta\le 1/2$.
     \begin{itemize}
         \item[(i)] $\mathbb{E} ( R_i^{2m} U_{i, j}^{2})$ = $\zeta_{2m}p^{- 1}\sigma_{jj}$ + $O( p^{m- 1- \delta / 2})$ for $j$ = $1, \ldots , p$ and $\mathbb{E} ( R_i^{2m} U_{i, j}U_{i, l})$ = $\zeta_{2m}p^{- 1}\sigma _{j\ell }$ + $O( p^{m- 1- \delta / 2})$ for $1\leqslant j\neq\ell\leqslant p.$
         \item[(ii)]if $\log p=o(n^{1/3})$ $$\left\|n^{-1/2}\sum_{i=1}^n\zeta_{m-1}^{-1}R_i^m\bm{U}_i\right\|_\infty=O_p\{\log^{1/2}(np)\}\quad\text{and}\quad\left\|n^{-1}\sum_{i=1}^n(\zeta_{m-1}^{-1}R_i^m\bm{U}_i)^2\right\|_\infty=O_p(1)\:.$$
     \end{itemize}
\end{lemma}
\begin{lemma}\label{covergence of zeta}
    Suppose the Assumptions in Lemma \ref{Convergence of D} hold, then $\hat{\zeta}_m\overset{p}{\rightarrow}\zeta_m$.
\end{lemma}
\begin{lemma}\label{bound of wU}
Suppose the Assumptions in Lemma \ref{777} hold with $a_0(p)\asymp p^{1-\delta}$ for some positive constant $\delta\le 1/2$ Then, if $\log p=o(n^{1/3})$
\begin{equation}\label{l8}
    \begin{aligned}&(i)\left\|n^{-1}\sum_{i=1}^{n}\zeta_{m-1}^{-1}w(\hat{R}_i)\hat{\boldsymbol{U}}_{i}\right\|_{\infty}=O_{p}\left\{n^{-1/2}\log^{1/2}(np)\right\},\\&(ii)\left\|\zeta_{m-1}^{-1}n^{-1}\sum_{i=1}^{n}\delta_{1i}w(\hat{R}_i)\hat{\boldsymbol{U}}_{i}\right\|_{\infty}=O_{p}(n^{-1}).\end{aligned}
\end{equation}
where $\delta_{i1}$ is defined in the proof in Lemma \ref{Bahadur represengtation}.
\end{lemma}
\begin{lemma}[Nazarov’s inequality]\label{12}
    Let $\bm{Y}_0=(Y_{0,1},Y_{0,2},\cdots,Y_{0,p})^\top$ be a centered Gaussian random vector in $\mathbb{R}^p$ and $\mathbb{E}(Y_{0,j}^2)\ge b$ for all $j=1,2,\cdots,p$ and some constant $b>0$, then for every $y\in \mathbb{R}^p$ and $a >0$,
    $$\mathbb{P}(\boldsymbol{Y}_{0}\leq y+a)-\mathbb{P}(\boldsymbol{Y}_{0}\leq y)\lesssim a\log^{1/2}(p).$$
\end{lemma}
\begin{lemma}[Theorem 2 in \cite{C15}]\label{13}
    Let $\bm{X}=(X_1,\cdots,X_p)^\top$ and $\bm{Y}=(Y_1,\cdots,Y_p)^\top$ be centered Gaussian random vectors in $\mathbb{R}^p$ with covariance matrices $\mathbf{\Sigma}^X=(\sigma^X_{jk})_{1\le j,k,\le p}$ and $\mathbf{\Sigma}^Y=(\sigma^Y_{jk})_{1\le j,k,\le p}$, respectively, In terms of $p$
    \begin{equation*}
        \Delta:=\max_{1\leq j,k\leq p}\left|\sigma_{jk}^X-\sigma_{jk}^Y\right|,anda_p:=\operatorname{E}\left[\max_{1\leq j\leq p}\left(Y_j/\sigma_{jj}^Y\right)\right].
    \end{equation*}
    Suppose that $p\ge 2$ and $\sigma_{jj}^Y>0$ for all $1\le j\le p$. Then
    \begin{align*}
\sup_{x\in\mathbb{R}}\left|\mathrm{P}\left(\max_{1\leq j\leq p}X_j\leq x\right)-\mathrm{P}\left(\max_{1\leq j\leq p}Y_j\leq x\right)\right| \\
\leq C\Delta^{1/3}\left\{\left(1\lor a_p^2\lor\log(1/\Delta)\right\}^{1/3}\log^{1/3}p,\right.
\end{align*}
where $C>0$ depends only on $\min_{1\le j\le p}\sigma_{jj}^Y$ and $\max_{1\le j\le p}\sigma_{jj}^Y$ (the right side is understood to be $0$ when $\Delta=0$). Moreover, in the worst case, $a_p\le\sqrt{2\log p}$, so that
\begin{equation*}
    \sup_{x\in\mathbb{R}}\left|\operatorname{P}\left(\max_{1\leq j\leq p}X_j\leq x\right)-\operatorname{P}\left(\max_{1\leq j\leq p}Y_j\leq x\right)\right|\leq C^{\prime}\Delta^{1/3}\{1\lor\log(p/\Delta)\}^{2/3},
\end{equation*}
where as before $C'\ge 0$ depends only on $\min_{1\le j\le p}\sigma_{jj}^Y$ and $\max_{1\le j\le p}\sigma_{jj}^Y$ .
\end{lemma}
\begin{lemma}[Theorem 1.2.11 in \cite{Mui}]\label{14}Let $X\sim N(\bm{\mu},\mathbf{\Sigma})$ with invertible $\mathbf{\Sigma}$, and partition $\bm{X}$, $\bm{\mu}$ and $\mathbf{\Sigma}$ as
$$\boldsymbol{X}=
\begin{pmatrix}
\boldsymbol{X}_1 \\
\boldsymbol{X}_2
\end{pmatrix},\quad\boldsymbol{\mu}=
\begin{pmatrix}
\boldsymbol{\mu}_1 \\
\boldsymbol{\mu}_2
\end{pmatrix},\quad\boldsymbol{\Sigma}=
\begin{pmatrix}
\boldsymbol{\Sigma}_{11} & \boldsymbol{\Sigma}_{12} \\
\boldsymbol{\Sigma}_{21} & \boldsymbol{\Sigma}_{22}
\end{pmatrix}.$$
Then $\boldsymbol{X}_{2}-\boldsymbol{\Sigma}_{21}\boldsymbol{\Sigma}_{11}^{-1}\boldsymbol{X}_{1}\sim N\left(\boldsymbol{\mu}_{2}-\boldsymbol{\Sigma}_{21}\boldsymbol{\Sigma}_{11}^{-1}\boldsymbol{\mu}_{1},\boldsymbol{\Sigma}_{22\cdot1}\right)$ and is independent of $\bm{X}_1$, where $\boldsymbol{\Sigma}_{22\cdot1}=\boldsymbol\Sigma_{22}-\boldsymbol\Sigma_{21}\boldsymbol\Sigma_{11}^{-1}\boldsymbol\Sigma_{12}.$
\end{lemma}
\subsection{Proof of main lemmas}
\subsubsection{Proof of Lemma \ref{GA}}
\begin{proof}
     Let $L_{n, p}= n^{- 1/ 4}\log ^{1/ 2}( np) + p^{- ( 1/ 6\wedge \delta / 2) }\log ^{1/ 2}( np) + n^{- 1/ 2}( \log p) ^{1/ 2}\log ^{1/ 2}( np) .$Then for any sequence $\eta_m\to\infty$ and any $t\in\mathbb{R}^p$,
$$\begin{aligned}\mathbb{P}(n^{1/2}\hat{\mathbf{D}}^{-1/2}(\hat{\boldsymbol{\theta}}-\boldsymbol{\theta})\leq t)&=\mathbb{P}(n^{-1/2}\zeta_{m-1}^{-1}\sum_{i=1}^{n}R_i^m\bm{U}_{i}+C_{n}\leq t)\\&\leq\mathbb{P}(n^{-1/2}\zeta_{m-1}^{-1}\sum_{i=1}^{n}R_i^m\bm{U}_{i}\leq t+\eta_{n}L_{n,p})+\mathbb{P}(\|C_{n}\|_{\infty}>\eta_{n}L_{n,p}).\end{aligned}$$
According to Lemma A4. in \cite{C23} and Cauchy inequality $\mathbb{E}\{(\zeta_{m-1}^{-1}R_i^m\boldsymbol{U}_{i,j})^{4}\}\lesssim\bar{M}^{2}$ and $\mathbb{E}\{(\zeta_{m-1}^{-1}R_i^m\boldsymbol{U}_{i,j})^{2}\}\gtrsim\underline{m}$ for all $i=1,2,\cdots,n$, $j=1,2,\cdots,p$, and the Gaussian approximation for independent partial sums in \cite{K21}, let $G\sim N\left(0,\zeta_{1}^{-2}\Sigma_{u}\right)$ with $\Sigma_w=\mathbb{E}\left(w^2(R_1)\boldsymbol{U}_{1}\boldsymbol{U}_{1}^{\top}\right)$, we have
$$\begin{aligned}\mathbb{P}(n^{1/2}\zeta_{1}^{-1}\sum_{i=1}^{n}R_i^mU_{i}\leq t+\eta_{n}L_{n,p})&\leq\mathbb{P}(G\leq t+\eta_{n}L_{n,p})+O(\{n^{-1}\log^{5}(np)\}^{1/6})\\&\leq\mathbb{P}(G\leq t)+O\{\eta_{n}L_{n,p}\log^{1/2}(p)\}+O(\{n^{-1}\log^{5}(np)\}^{1/6}).\end{aligned}$$
where the second inequality holds from Nazarov's inequality in Lemma \ref{12} Thus,
$$\begin{aligned}\mathbb{P}(n^{1/2}\hat{\mathbf{D}}^{-1/2}(\hat{\boldsymbol{\theta}}-\boldsymbol{\theta})\leq t)\leq&\mathbb{P}(\boldsymbol{G}\leq t)+O\{\eta_{n}L_{n,p}\log^{1/2}(p)\}+O(\{n^{-1}\log^{5}(np)\}^{1/6})\\&+\mathbb{P}(|C_{n}|_{\infty}>\eta_{n}l_{n,p}).\end{aligned}$$
On the other hand, we have
$$\mathbb{P}(n^{1/2}\hat{\mathbf{D}}^{-1/2}(\hat{\boldsymbol{\theta}}-\boldsymbol{\theta})\leq t)\geq\mathbb{P}(\boldsymbol{G}\leq t)-O\{\eta_{n}L_{n,p}\log^{1/2}(p)\}-O(\{n^{-1}\log^{5}(np)\}^{1/6})\\-\mathbb{P}(\|C_{n}\|_{\infty}>\eta_{n}l_{n,p}).$$
where $\mathbb{P}(\|C_n\|_\infty>\eta_nl_{n,p})\to0$ as $n\to\infty$ by Theorem \ref{Bahadur represengtation}
Then we have that, if $\log p=o(n^{1/5})$ and $\log n=o(p^{1/3}\wedge\delta)$,
$$\sup\limits_{t\in\mathbb{R}^p}|\mathbb{P}(n^{1/2}\hat{\mathbf{D}}^{-1/2}(\hat{\boldsymbol{\theta}}-\boldsymbol{\theta})\leq t)-\mathbb{P}(\boldsymbol{G}\leq t)|\to0.$$
Further,
$$\rho_{n}(\mathcal{A}^{re})=\sup_{A\in\mathcal{A}^{re}}|\mathbb{P}(n^{1/2}\hat{\mathbf{D}}^{-1/2}(\hat{\boldsymbol{\theta}}-\boldsymbol{\theta})\in A)-\mathbb{P}(\boldsymbol{G}\in A)|\to0,$$
by the Corollary 5.1 in \cite{C17}
\end{proof}
\subsubsection{Proof of Lemma \ref{VA}}
\begin{proof}
$\mathbb{E} Z_j^2= \zeta_{m-1}^{- 2}\mathbb{E}( R_i^{2m}) ^{- 1}\leq \bar{B}$ by Assumption \ref{order} and $\mathbb{E}[\max_{1\leq j\leq p}Z_j]\asymp(\sqrt{\log p+\log\log p})$ by Theorem ${2}$ in \cite{F22a}. Let $\Delta_0=\max_{1\leq j,k\leq p}|p(\mathbb{E}w^2(R_1)\bm{U}_1\bm{U}_1^\top)_{j,k}-\mathbf{R}_{j,k}|$, by Lemma \ref{Qh}
$$\Delta_{0}=\max_{1\leq j,k\leq p}|p(\mathbb{E}w^2(R_1)\bm{U}_{1}\bm{U}_{1}^{\top})_{j,k}-\mathbf{R}_{j,k}|=O(p^{-\delta/2}).$$
According to Lemma \ref{13},we get
$$\sup\limits_{t\in\mathbb{R}}|\mathbb{P}\left(\|\boldsymbol{Z}\|_{\infty}\leqslant t\right)-\mathbb{P}\left(\|\boldsymbol{G}\|_{\infty}\leqslant t\right)|\leqslant C'n^{-1/3}\left(1\vee\log\left(np\right)\right)^{2/3}\to0.$$
\end{proof}
\subsubsection{Proof of Lemma \ref{order of varU}}
\begin{proof}
    By Cauchy inequality and Assumption \ref{struction of W} we have
$$\begin{aligned}&\mathbb{E}U(\boldsymbol{W}_{i})_{l}^{2}U(\boldsymbol{W}_{i})_{k}^{2}\leq\frac{1}{p^{2}}\mathbb{E}\sum_{s=1}^{p}\sum_{t=1}^{p}U(\boldsymbol{W}_{i})_{s}^{2}U(\boldsymbol{W}_{i})_{t}^{2}=p^{-2}\\&\mathbb{E}U(\boldsymbol{W}_{i})_{l}^{4}\leq\frac{1}{p}\mathbb{E}\sum_{s=1}^{p}U(\boldsymbol{W}_{i})_{s}^{4}\leq\frac{1}{p}\mathbb{E}\sum_{s=1}^{p}\sum_{t=1}^{p}U(\boldsymbol{W}_{i})_{s}^{2}U(\boldsymbol{W}_{i})_{t}^{2}=p^{-1},\end{aligned}$$
and
$$\mathbb{E}\left(U(\boldsymbol{W}_i)_lU(\boldsymbol{W}_i)_kU(\boldsymbol{W}_i)_sU(\boldsymbol{W}_i)_t\right)\leq\sqrt{E\left(U(\boldsymbol{W}_i)_l^2U(\boldsymbol{W}_i)_k^2\right)E\left(U(\boldsymbol{W}_i)_s^2U(\boldsymbol{W}_i)_t^2\right)}.$$
By the Cauchy inequality,
$$\sum_{i,k,s,t}a_{lk}a_{st}\leq\sqrt{\sum_{l,k}a_{lk}^{2}\sum_{s,t}a_{st}^{2}}\leq\sqrt{\sum_{l,k}^{p}a_{lk}^{2}\sum_{s,t}^{p}a_{st}^{2}}=\mathrm{tr}(\mathbf{M}^{\mathsf{T}}\mathbf{M}).$$
Thus, we get
$$\begin{aligned}&E\left(\left(U(\boldsymbol{W}_{i})^{\top}\mathbf{M}\mathbf{U}(\mathbf{W}_{i})\right)^{2}\right)\\&=\sum_{l\neq k=1}^{p}\sum_{s\neq l=1}^{p}a_{lk}a_{sl}\mathbb{E}\left(U(\boldsymbol{W}_{i})_{l}U(\boldsymbol{W}_{i})_{k}U(\boldsymbol{W}_{i})_{s}U(\boldsymbol{W}_{i})_{t}\right)+\sum_{l=1}^{p}\sum_{s=1}^{p}a_{ll}a_{ss}\mathbb{E}\left(U(\boldsymbol{W}_{i})_{l}^{2}U(\boldsymbol{W}_{i})_{s}^{2}\right)\\&\leq p^{-2}\frac{p^{4}-p^{2}}{p^{4}}\mathrm{tr}(\mathbf{M}^{\top}\mathbf{M})+p^{-1}\frac{p^{2}}{p^{4}}\mathrm{tr}(\mathbf{M}^{\top}\mathbf{M})=O(p^{-2}\mathrm{tr}(\mathbf{M}^{\top}\mathbf{M})).\end{aligned}$$
\end{proof}
\subsubsection{Proof of Lemma \ref{Convergence of D}}
\begin{proof}
This proof is based on Lemma A.2. in  and extends it to the weighted case.

Denote $\bm{\eta}=(\zeta_m^{-1}\bm{\theta}^\top ,d_1,d_2,\dots,d_p)^\top$ and $\hat{\bm{\eta}}$ as the estimated version. By first-order Taylor expansion, we have
\begin{align}\label{111}
w(\|\bm{\epsilon}_i\|)U(\bm{\epsilon}_i)&=\frac{w(R_i)\mathbf{D}^{-1/2}\mathbf{\Sigma}^{1/2}U(\bm{W}_i)}{(1+U(\bm{W}_i)^\top(\mathbf{R}-\mathbf{I}_p)U(\bm{W}_i))^{1/2}}\\
&=w(R_i)\mathbf{R}^{-1/2}\mathbf{\Sigma}^{1/2}U(\bm{W}_i)+C_1w(R_i)U(\bm{W}_i)^\top(\mathbf{R}-\mathbf{I}_p)U(\bm{W}_i)\mathbf{D}^{-1/2}\mathbf{\Sigma}^{1/2}U(\bm{W}_i),
\end{align}
where $C_1$ is a bounded random variable between $-0.5$ and $-0.5(1+U(\bm{W}_i)^\top(\mathbf{R}-\mathbf{I}_p)U(\bm{W}_i))^{-3/2}$. By Cauchy inequality, Lemma \ref{order of varU} and , we have
\begin{align*}
    \mathbb{E}(w(R_i)U(\bm{\epsilon}_i)_j)&\le C_2\{\mathbb{E}(w^2(R_i))\mathbb{E}((U(\bm{W}_i)^\top(\mathbf{R}-\mathbf{I}_p)U(\bm{W}_i))^2)\mathbb{E}(\mathbf{D}^{-1/2}\mathbf{\Sigma}^{1/2}U(\bm{W}_i))_j^2\}^{1/2}\\
    &=O(\zeta_m p^{-1}\sqrt{\mathrm{tr}(\mathbf{R}^2)-p})=o(\zeta_mn^{-1/2}).
\end{align*}
Similarly, we can show that
\begin{equation*}
    \mathbb{E}\left( \mathrm{diag}\{U(\mathbf{D}^{-1/2}(\bm{X}_i-\bm{\theta}))U(\mathbf{D}^{-1/2}(\bm{X}_i-\bm{\theta}))^\top\}-\frac{1}{p}\mathbf{I}_p \right)\le O(n^{-1/2}),
\end{equation*}
by first-order Taylor expansion for $U(\mathbf{D}^{-1/2}(X_i-\theta))U(\mathbf{D}^{-1/2}(X_i-\theta))^\top$, Cauchy inequality and Lemma  The above two equations define the functional equation for each component of $\eta$,

\begin{equation}\label{T}
    T_j(F,\eta_j)=o_p(n^{-1/2}),
\end{equation}
where $F$ represent the distribution of $X,\boldsymbol{\eta}=(\eta_1,\cdots,\eta_{2p}).$ Similar with \cite{HR}, the linearisation of this equation shows
$$n^{1/2}\left(\hat{\eta}_{j}-\eta_{j}\right)=-\mathbf{H}_{j}^{-1}n^{1/2}\left(T_{j}(F_{n},\eta_{j})-T_{j}(F,\eta_{j})\right)+o_{p}(1),$$
where $F_n$ represents the empirical distribution function based on $\bm{X}_1,\bm{X}_2,\cdots,\bm{X}_n,\mathbf{H}_j$ is the corresponding Hessian matrix of the functional defined in Equation\ref{T}and
$$T(F_n,\boldsymbol{\eta})=\left(n^{-1}\zeta_m^{-1}\sum_{j=1}^nw(R_i)U(\bm{\epsilon}_i)^\top,\mathrm{vec}(\mathrm{diag}(n^{-1}U(\bm{\epsilon}_i)U(\bm{\epsilon}_i)^\top-\frac{1}{p}\mathbf{I}_p))\right).$$
Thus, for each $\hat{d}_j$ we have
$$\sqrt{n}(\hat{d}_j-d_j)\stackrel{d}{\to}N(0,\sigma_{d,j}^2).$$
where $\sigma_{d,j}^2$ is the corresponding asymptotic variance. Define $\sigma_{d,max}=\max_{1\leq j\leq p}\sigma_{d,j}.$ As
$$\begin{aligned}&\mathbb{P}(\max_{j=1,2,\cdots,p}(\hat{d}_{j}-d_{j})>\sqrt{2}\sigma_{d,max}n^{-1/2}(\log p)^{1/2})\\&\leq\sum_{j=1}^{p}\mathbb{P}(\sqrt{n}(\hat{d}_{j}-d_{j})>\sqrt{2}\sigma_{d,max}(\log p)^{1/2})\\&=\sum_{j=1}^{p}(1-\Phi(\sqrt{2}\sigma_{d,max}\sigma_{d,j}^{-1}(\log p)^{1/2}))\leq p(1-\Phi((2\log p)^{1/2}))\\&\leq\frac{p}{\sqrt{2\pi}\log p}\exp(-\log p)=(4\pi)^{-1/2}(\log p)^{-1/2}\to0,\end{aligned}$$
which means that $\max_{j=1,2,\dots,p}(\hat{d}_j-d_j)=O_p(n^{-1/2}(\log p)^{1/2}).$
\end{proof}
\subsubsection{Proof of Lemma \ref{Qh}}
\begin{proof}
Denote $\mathbf{\Omega}=\mathbf{D}^{-1/2}\mathbf{\Gamma}$ and $\hat{\mathbf{\Omega}}=\hat{\mathbf{D}}^{-1/2}\mathbf{\Gamma}$. Set $\bm{\Omega}_i^\top$ ,$\bm{\hat{\Omega}}_i^\top$ and $\bm{\Gamma}_i^\top$ be the $i$th row of $\mathbf{\Omega}$ ,$\mathbf{\hat{\Omega}}$ and
$\mathbf{\Gamma}$ respectively.
\begin{align*}
    \hat{\mathbf{Q}}_{jl}(s)&=\frac{1}{n}\sum_{i=1}^n \hat{R}_i^s\hat{\bm{U}}_{ij}\hat{\bm{U}}_{il}^\top\\
    &=\frac{1}{n}\sum_{i=1}^n{\hat{R}_i^{s-2}}v_i^2\hat{\bm{\Omega}}_j^\top\bm{W}_i\hat{\bm{\Omega}}_l^\top\bm{W}_i\\
    &=\frac{1}{n}\sum_{i=1}^n\left\|\hat{\mathbf{\Omega}}\bm{W}_i\right\|^{s-2}v_i^s\hat{d}_j^{-1}\hat{d}_l^{-1}{\bm{\Gamma}}_j^\top\bm{W}_i{\bm{\Gamma}}_l^\top\bm{W}_i\\
    &=A_1+A_2+A_3,
\end{align*}
    where $A_1$, $A_2$ and $A_3$ are defined as follows
\begin{align*}
    A_1&=\frac{1}{n}\sum_{i=1}^n\left\{\left\|\hat{\mathbf{\Omega}}\bm{W}_i\right\|^{s-2}-\left\|{\mathbf{\Omega}}\bm{W}_i\right\|^{s-2}\right\}v_i^s\hat{d}_j^{-1}\hat{d}_l^{-1}{\bm{\Gamma}}_j^\top\bm{W}_i{\bm{\Gamma}}_l^\top\bm{W}_i;\\
   A_2&=\frac{1}{n}\sum_{i=1}^n\left\{\left\|{\mathbf{\Omega}}\bm{W}_i\right\|^{s-2}-p^{(s-2)/2}\right\}v_i^s\hat{d}_j^{-1}\hat{d}_l^{-1}{\bm{\Gamma}}_j^\top\bm{W}_i{\bm{\Gamma}}_l^\top\bm{W}_i;\\
   A_3&=\frac{1}{n}\sum_{i=1}^np^{(s-2)/2}v_i^s\hat{d}_j^{-1}\hat{d}_l^{-1}{\bm{\Gamma}}_j^\top\bm{W}_i{\bm{\Gamma}}_l^\top\bm{W}_i.
\end{align*}
From the proof of Lemma 8 in ref, when the Lemma \ref{Convergence of D} holds, we have
\begin{align*}
&\|\hat{\mathbf{D}}^{-1/2}\mathbf{\Gamma}\boldsymbol{W}_{i}\|^{2}\\
=&\|(\hat{\mathbf{D}}^{-1/2}\mathbf{D}^{1/2}-\mathbf{I}_{p})\mathbf{D}^{-1/2}\mathbf{\Gamma}\boldsymbol{W}_{i}+\mathbf{D}^{-1/2}\mathbf{\Gamma}\boldsymbol{W}_{i}\|^{2}\\
=&\|\mathbf{D}^{-1/2}\mathbf{\Gamma}\boldsymbol{W}_{i}\|^{2}+\|(\hat{\mathbf{D}}^{-1/2}\mathbf{D}^{1/2}-\mathbf{I}_{p})\mathbf{D}^{-1/2}\mathbf{\Gamma}\boldsymbol{W}_{i}\|^{2}+\boldsymbol{W}_{i}\mathbf{\Gamma}^{\top}\mathbf{D}^{-1/2}(\hat{\mathbf{D}}^{-1/2}\mathbf{D}^{1/2}\mathbf{I}_{p})\mathbf{D}^{-1/2}\mathbf{\Gamma}\boldsymbol{W}_{i}\\
\leq&\|\mathbf{D}^{-1/2}\mathbf{\Gamma}\boldsymbol{W}_{i}\|^{2}\left(1+\max_{i=1,2,\cdots,p}(\frac{d_{i}}{\hat{d}_{i}}-1)^{2}+\max_{i=1,2,\cdots,p}(\frac{d_{i}}{\hat{d}_{i}}-1)\right)\\
:=&\|\mathbf{D}^{-1/2}\mathbf{\Gamma}\boldsymbol{W}_{i}\|^{2}\left(1+H\right),
\end{align*}
where $H:=\max_i(\frac{d_i}{\hat{d}_i}-1)^2+\max_i(\frac{d_i}{\hat{d}_i}-1)=O_p(n^{-1/2}(\log p)^{1/2}).$
For any $k$,
\begin{align}\label{H}
\|\hat{\mathbf{D}}^{-1/2}\mathbf{\Gamma}\boldsymbol{W}_{i}\|^{k}&=\|(\hat{\mathbf{D}}^{-1/2}\mathbf{D}^{1/2}-\mathbf{I}_{p})\mathbf{D}^{-1/2}\mathbf{\Gamma}\boldsymbol{W}_{i}+\mathbf{D}^{-1/2}\mathbf{\Gamma}\boldsymbol{W}_{i}\|^{k}\\&=\left\{\|(\hat{\mathbf{D}}^{-1/2}\mathbf{D}^{1/2}-\mathbf{I}_{p})\mathbf{D}^{-1/2}\mathbf{\Gamma}\boldsymbol{W}_{i}+\mathbf{D}^{-1/2}\mathbf{\Gamma}\boldsymbol{W}_{i}\|^{2}\right\}^{k/2}\\&\leq\|\mathbf{D}^{-1/2}\mathbf{\Gamma}\boldsymbol{W}_{i}\|^{k}\left(1+H\right)^{k/2}\\&:=\|\mathbf{D}^{-1/2}\mathbf{\Gamma}\boldsymbol{W}_{i}\|^{k}\left(1+H_{k}\right),
\end{align}
where $H_k$ is defined as $H_k=(1+H)^{k/2}-1=O_p(n^{-1/2}(\log p)^{1/2}).$

From the proof of Lemma A3 in \cite{C23} we can conclude
\begin{align*}
    &\mathbb{E}A_1\\
    =&\mathbb{E}\left\{\frac{1}{n}\sum_{i=1}^n\left\{\left\|\hat{\mathbf{\Omega}}\bm{W}_i\right\|^{s-2}-\left\|{\mathbf{\Omega}}\bm{W}_i\right\|^{s-2}\right\}v_i^s\hat{d}_j^{-1}\hat{d}_l^{-1}{\bm{\Gamma}}_j^\top\bm{W}_i{\bm{\Gamma}}_l^\top\bm{W}_i\right\}\\
    =&\mathbb{E}\left\{\frac{1}{n}\sum_{i=1}^n\left\{\left\|{\mathbf{\Omega}}\bm{W}_i\right\|^{s-2}H_{s-2}\right\}v_i^s\hat{d}_j^{-1}\hat{d}_l^{-1}{\bm{\Gamma}}_j^\top\bm{W}_i{\bm{\Gamma}}_l^\top\bm{W}_i\right\}\\
    =&\mathbb{E}\left\{(A_2+A_3)H_{s-2}\right\}.
\end{align*}
Next, we analyze $A_2$ and $A_3$. From the proof of Lemma A3 in \cite{C23} we can conclude
\begin{align*}
    d_j^{-1}d_l^{-1}\hat{d_j}\hat{d_l}A_2&=\frac{1}{n}\sum_{i=1}^n\left\{\left\|{\mathbf{\Omega}}\bm{W}_i\right\|^{s-2}-p^{(s-2)/2}\right\}v_i^s{d}_j^{-1}{d}_l^{-1}{\bm{\Gamma}}_j^\top\bm{W}_i{\bm{\Gamma}}_l^\top\bm{W}_i\\
    &=O_p(p^{(s-2)/2-\delta/2}\zeta_s p^{-s/2})\\
    &=O_p(\zeta_s p^{-1-\delta/2}),
\end{align*}
and
\begin{align*}
    d_j^{-1}d_l^{-1}\hat{d_j}\hat{d_l}A_3&=\frac{1}{n}\sum_{i=1}^np^{(s-2)/2}v_i^s{d}_j^{-1}{d}_l^{-1}{\bm{\Gamma}}_j^\top\bm{W}_i{\bm{\Gamma}}_l^\top\bm{W}_i\\
    &\lesssim \zeta_s p^{-1}|\sigma_{jl}|+O_p\left( \zeta_s n^{-1/2}p^{-1}+\zeta_s p^{-7/6} \right).
\end{align*}
It follows that,
\begin{equation*}
     d_j^{-1}d_l^{-1}\hat{d_j}\hat{d_l}|\hat{\mathbf{Q}}_{jl}|\lesssim\left(  \zeta_s p^{-1}|\sigma_{jl}|+O_p\left( \zeta_s n^{-1/2}p^{-1}+\zeta_s p^{-7/6}+\zeta_s p^{-1-\delta/2} \right)\right)(1+H_{s-2}).
\end{equation*}
Recall lemma \ref{Convergence of D}, we have
\begin{equation*}
    |\hat{\mathbf{Q}}_{jl}|\lesssim\left(  \zeta_s p^{-1}|\sigma_{jl}|+O_p\left( \zeta_s n^{-1/2}p^{-1}+\zeta_s p^{-7/6}+\zeta_s p^{-1-\delta/2} \right)\right)(1+O_p(n^{-1/2}(\log p)^{1/2}))^2.
\end{equation*}
Thus,
\begin{equation*}
     |\hat{\mathbf{Q}}_{jl}|\lesssim \zeta_s p^{-1}|\sigma_{jl}|+O_p\left( \zeta_s n^{-1/2}p^{-1}+\zeta_s p^{-7/6}+\zeta_s p^{-1-\delta/2}+\zeta_{s}n^{-1/2}(\log p)^{1/2}(p^{-5/2}+p^{-1-\delta/2} )\right).
\end{equation*}
\end{proof}
\subsubsection{Proof of Lemma \ref{777}}
\begin{proof}
    (i) By Lemma 6 with $s=2m$. we can get it easily.

    (ii)By Assumption \ref{struction of W} and \ref{order}, we can get $\zeta_{m-1}^{-1}R_i^m\bm{U}_1,\ldots,\zeta_{m-1}^{-1}R_i^m\bm{U}_n$ are i.i.d. $p$-dimensional random vectors satisfies $\|\zeta_{m-1}^{-1}R_i^m\bm{U}_{i,j}\|_{\psi_{\alpha}}\lesssim c\bar{B}$ for all $i=1,\ldots,n$ and $j=1,\ldots,p.$ By Lemma 2.2.2 of van der Vaart ,
$$\left\|\max\limits_{1\leqslant i\leqslant n}\max\limits_{1\leqslant j\leqslant p}|\zeta_{m-1}^{-1}R_i^m\bm{U}_{i,j}|\right\|_{\psi_{\alpha}}\leqslant\log^{1/\alpha}(np)\:.$$
Similar to the proof of lemma 6 , we can show that
$$\begin{aligned}\mathbb{E}\{(\zeta_{m-1}^{-1}R_i^m\bm{U}_{i,j})^{2}\}&=\quad\zeta_{m-1}^{-2}\mathbb{E}\{v_i^{2m}\|\bm{\Gamma} \bm{W}_{i}\|^{2m-2}(\bm{\Gamma}_{j}\bm{W}_{i})^{2}\mathbb{I}(\mathcal{A}_{1i})\}\\
&+\zeta_{m-1}^{-2}\mathbb{E}\{v_i^{2m}\|\bm{\Gamma} \bm{W}_{i}\|^{2m-2}(\bm{\Gamma}_{j}\bm{W}_{i})^{2}\mathbb{I}(\mathcal{A}_{1i}^{c})\}\\
&\lesssim \zeta_{m-1}^{-2}\zeta_{4m}^{1/2}/p^{m}\cdot p^{m-1}d_{j}+\zeta_{m-1}^{-2}c_{1}\exp\{-c_{2}p^{\delta/(4+4\alpha)}\}
&\lesssim d_{j}
\:.\end{aligned}$$
It follows that
$$\begin{array}{rcl}\max\limits_{1\leqslant j\leqslant p}\sum\limits_{i=1}^n\mathbb{E}\{(\zeta_{m-1}^{-1}R_i^m\bm{U}_{i,j})^2\}&\leqslant n\max\limits_{1\leqslant j\leqslant p}d_{j}\lesssim n\:,\end{array}$$
Applying Lemma E.1 of \cite{C17}, it holds that with $\alpha\geqslant1$
and $n^{-1/2}\log^{3/2}(np)\lesssim1$,
$$\begin{aligned}\mathbb{E}\left(\left|n^{-1/2}\sum_{i=1}^{n}\zeta_{m-1}^{-1}R_i^m\bm{U}_{i}\right|_{\infty}\right)&\leqslant\quad n^{-1/2}\{n^{1/2}\log^{1/2}(p)+\log^{1/\alpha}(np)\log(p)\}\\&\lesssim\quad\log^{1/2}(np)\:.\end{aligned}$$
From the properties of the $\psi_{\alpha}$ norm, it holds that
$$\left\|\max_{1\leqslant i\leqslant n,1\leqslant j\leqslant p}|\zeta_{m-1}^{-1}R_i^m\bm{U}_{i,j}|^2\right\|_{\psi_{\alpha/2}}\leqslant\log^2(np).$$
According to Lemma E.3 of \cite{C17}, we have that
$$\mathbb{E}\left(\left|n^{-1}\sum_{i=1}^n(\zeta_{m-1}^{-1}R_i^m\bm{U}_i)^2\right|_\infty\right)\:\leqslant\:n^{-1}\{n+\log^2(np)\log(p)\}\leqslant1\:.$$
We finish the proof of this lemma.
\end{proof}
\subsubsection{Proof of Lemma \ref{covergence of zeta}}
\begin{proof}
    Denote $\hat{\boldsymbol{\bm{\mu}}}=\hat{\boldsymbol{\theta}}-\boldsymbol{\theta}.$
$$\begin{aligned}\|\hat{\mathbf{D}}^{-1/2}(\boldsymbol{X}_{i}-\hat{\boldsymbol{\theta}})\|&=\|\mathbf{D}^{-1/2}(\boldsymbol{X}_{i}-\boldsymbol{\theta})\|(1+R_{i}^{-2}\|(\hat{\mathbf{D}}^{-1/2}-\mathbf{D}^{-1/2})(\boldsymbol{X}_{i}-\boldsymbol{\theta})\|^{2}\\&+R_{i}^{-2}\|\hat{\mathbf{D}}^{-1/2}\hat{\boldsymbol{\bm{\mu}}}\|^{2}+2R_{i}^{-2}\boldsymbol{U}_{i}^{\top}(\hat{\mathbf{D}}^{-1/2}-\mathbf{D}^{-1/2})\mathbf{D}^{1/2}\boldsymbol{U}_{i})\\&-2R_{i}^{-1}\boldsymbol{U}_{i}^{\top}\hat{\mathbf{D}}^{-1/2}\hat{\boldsymbol{\bm{\mu}}}-2R_{i}^{-1}\boldsymbol{U}_{i}\mathbf{D}^{1/2}(\hat{\mathbf{D}}^{-1/2}-\mathbf{D}^{-1/2})\hat{\mathbf{D}}^{-1/2}\hat{\boldsymbol{\bm{\mu}}})^{1/2}.\end{aligned}$$
According to the proof and conclusion in Lemma \ref{Convergence of D} we can show that $R_i^{-2}\|(\hat{\mathbf{D}}^{-1/2}\boldsymbol{-}\mathbf{D}^{-1/2})(\boldsymbol{X}_i\boldsymbol{-}$
$\boldsymbol{\theta})\|^{2}=O_{p}\left((\log p/n)^{1/2}\right)=o_{p}(1)$ and $R_i^{-2}\|\hat{\mathbf{D}}^{-1/2}\hat{\boldsymbol{\bm{\mu}}}\|^{2}=O_{p}(n^{-1})=o_{p}(1)$ and by the Cauchy
inequality, the other parts are also $o_p(1).$ So,
$$n^{-1}\sum_{i=1}^n\left\|\hat{\mathbf{D}}^{-1/2}\left(\boldsymbol{X}_i-\hat{\boldsymbol{\theta}}\right)\right\|^{m}=\left(n^{-1}\sum_{i=1}^n\left\|\mathbf{D}^{-1/2}\left(\boldsymbol{X}_i-\boldsymbol{\theta}\right)\right\|^{m}\right)\left(1+o_p(1)\right).$$
Obviously,  $\mathbb{E}\left ( n^{- 1}\sum _{i= 1}^nR_i^{m}\right ) = \zeta _m$ and  $\mathrm{Var}\left ( n^{- 1}\zeta _m^{- 1}\sum _{i= 1}^nR_i^{m}\right ) = O\left ( n^{- 1}\right ) .$ Finally,  the proof
is completed.
\end{proof}
\subsubsection{Proof of Lemma \ref{bound of wU}}
\begin{proof}
For any $j\in \{ 1, 2, \cdots , p\} ,$
$$\begin{aligned}\hat{\boldsymbol{U}}_{ij}-\boldsymbol{U}_{ij}&=\frac{\|\mathbf{D}^{-1/2}X_{i}\|}{\|\hat{\mathbf{D}}^{-1/2}X_{i}\|}\cdot\frac{d_{j}}{\hat{d}_{j}}\boldsymbol{U}_{ij}-\boldsymbol{U}_{ij}\\&\leq(1+H)(1+H)\boldsymbol{U}_{ij}-\boldsymbol{U}_{ij}\\&=H_{u}\boldsymbol{U}_{ij},\end{aligned}$$
where $H_u=O_p(H^2+2H)=O_p(n^{-1/2}(\log p)^{1/2}).$ Thus, $\hat{\bm{U}}_i-\bm{U}_i=H_u\boldsymbol{U}_i.$

(i) By Equation \ref{bound of wU}, we have
\begin{align*}
&\left\|n^{-1}\sum_{i=1}^{n}\zeta_{m-1}^{-1}w(\hat{R}_i)\hat{\boldsymbol{U}}_{i}\right\|_{\infty}=\left\|n^{-1}\sum_{i=1}^{n}\zeta_{m-1}^{-1}(1+H_{u})w(\hat{R}_i)\boldsymbol{U}_{i}\right\|_{\infty}\\
\leq&|1+H_{u}|\cdot\left\|n^{-1}\sum_{i=1}^{n}\zeta_{m-1}^{-1}w(\hat{R}_i)\boldsymbol{U}_{i}\right\|_{\infty}=O_{p}\left\{n^{-1/2}\log^{1/2}(np)\right\}.
\end{align*}

(ii)Similarly
\begin{align*}
    &\left|\zeta_{m-1}^{-1}n^{-1}\sum_{i=1}^{n}\delta_{1i}w(\hat{R_i})\hat{\bm{U}}_{i}\right|_{\infty}\leq|1+H_{u}|\cdot\left\|\zeta_{m-1}^{-1}n^{-1}\sum_{i=1}^{n}\delta_{1i}w(\hat{R_i})\bm{U}_{i}\right\|_{\infty}\\
    \leq& O_{p}(n^{-1}(1+n^{-1/2}\log^{1/2}p))=O_{p}(n^{-1}).
\end{align*}
\end{proof}

\subsection{Proof of main results}
\subsubsection{Proof of Theorem \ref{Bahadur represengtation}}
\begin{proof}
    Note that given $\hat{\boldsymbol{D}}$, the estimator $\hat{\boldsymbol{\bm{\mu}}}$ is the minimizer of the following objective function
\begin{align*}
    L(\boldsymbol{\bm{\mu}})&=\sum_{i=1}^n\left\|\hat{\mathbf{D}}^{-1/2}(\mathbf{X}_{i}-\boldsymbol{\theta}-\boldsymbol{\bm{\mu}})\right\|^{m+1}\quad \mathrm{for}\;m\neq -1;\\
     L(\boldsymbol{\bm{\mu}})&=\sum_{i=1}^n\ln \left\|\hat{\mathbf{D}}^{-1/2}(\mathbf{X}_{i}-\boldsymbol{\theta}-\boldsymbol{\bm{\mu}})\right\|\qquad \mathrm{for}\;m=-1,
\end{align*}
and the estimator $\hat{\boldsymbol{\theta}}$ is equivalent to solve the equation
$$\sum_{i=1}^n w\left(\left\|\mathbf{{D}}^{-1/2}\left( \bm{X}_i-\bm{{\theta}} \right)\right\|\right) U(\hat{\mathbf{D}}^{-1/2}(\mathbf{X}_{i}-\boldsymbol{\theta}))=0.$$
By Assumption \ref{existence and uniqueness of HR}, we know the objective function $L(\bm{\mu})$ exists an unique local minimizer and it must be a global minimizer. Our goal is find $b_{n,p}$ such that $||\bm{\mu}||=O_p(b_{n,p})$. The existence of a $b_{n,p}^{-1}$-consistent local minimizer is implied by the fact that for an arbitrarily small $\epsilon>0$, there exist a sufficiently large constant $C$, which does no depend on $n$ or $p$, such that
\begin{equation}\label{consistent minimizer}
    \lim\inf_n\mathbb{P}\left( \inf_{\bm{u}\in\mathbb{R}^p,||\bm{u}||=C} L(b_{n,p}\bm{u})>L(\bm{0})\right)>1-\epsilon.
\end{equation}

Firstly, we prove Equation \ref{consistent minimizer} holds when $m\neq -1$ and $b_{n,p}=p^{1/2}n^{-1/2}$. Consider the expansion of $\left\|\hat{\mathbf{D}}^{-1/2}(\mathbf{X}_{i}-\boldsymbol{\theta}-b_{n,p}\bm{u})\right\|^{m+1}$:
\begin{align*}
    &\left\|\hat{\mathbf{D}}^{-1/2}(\mathbf{X}_{i}-\boldsymbol{\theta}-b_{n,p}\bm{u})\right\|^{m+1}\\
    =&\left\|\hat{\mathbf{D}}^{-1/2}(\mathbf{X}_{i}-\boldsymbol{\theta})\right\|^{m+1}\left(1-2b_{n,p}\hat{R}_i^{-1}\bm{u}^\top\hat{\mathbf{D}}^{-1/2}\hat{\bm{U}_i}+b_{n,p}^2\hat{R}^{-2}\bm{u}^\top\hat{\mathbf{D}}^{-1}\bm{u}\right)^{\frac{m+1}{2}}\\
    =&\hat{R}_i^{m+1}\Big( 1-(m+1)b_{n,p}\hat{R}_i^{-1}\bm{u}^\top \hat{\mathbf{D}}^{-1/2}\hat{\bm{U}}_i+ \frac{m+1}{2} b_{n,p}^2\hat{R}_i^{-2}\bm{u}^\top\hat{\mathbf{D}}^{-1}\bm{u}\\
    &+\frac{m^2-1}{2}b_{n,p}^2\hat{R}_i^{-2}\bm{u}^\top\hat{\mathbf{D}}^{-1/2}\hat{\bm{U}}_i^\top\hat{\bm{U}}_i\hat{\mathbf{D}}^{-1/2}\bm{u}+O_p(n^{-3/2})\Big)\\
    =&\hat{R}_i^{m+1}-(m+1)b_{n,p}\hat{R}_i^m\bm{u}^\top\hat{\mathbf{D}}^{-1/2}\hat{\bm{U}}_i\\
    &+\frac{m+1}{2}b_{n,p}^2\hat{R}_i^{m-1}\bm{u}\hat{\mathbf{D}}^{-1/2}\left( \mathbf{I}_p+(m-1)\hat{\bm{U}}_i\hat{\bm{U}}_i^\top \right)\hat{\mathbf{D}}^{-1/2}\bm{u}+O_p(\zeta_{m+1}n^{-3/2}).
\end{align*}
So, it can be easily seen
\begin{equation}\label{eq1}
\begin{aligned}
    &p^{-1/2}\zeta_{m}^{-1}\left( L(b_{n,p}\bm{u})-L(\bm{0}) \right)\\
    =&-n^{-1/2}\zeta_{m}^{-1}(m+1)\bm{u}^\top\hat{\mathbf{D}}^{-1/2}\sum_{i=1}^n\hat{R}_i^m\hat{\bm{U}}_i\\
    &+\frac{m+1}{2}p^{1/2}\zeta_m^{-1}\bm{u}\hat{\mathbf{D}}^{-1/2}\left(  \frac{1}{n}\sum_{i=1}^n\left( \hat{R}_i^{m-1}\mathbf{I}_p+(m-1)\hat{R}_i^{m-1}\hat{\bm{U}}_i\hat{\bm{U}}_i^\top  \right)\right)\hat{\mathbf{D}}^{-1/2}\bm{u}+O_p(n^{-1/2}).
\end{aligned}
\end{equation}
Under assumption \ref{order}, we can conclude $\mathbb{E}\left( \|n^{-1/2}\zeta_m^{-1}\sum_{i=1}^n\hat{R}^{m}_i\hat{\bm{U}}_i\|^2 \right)=O(1)$ and $\mathrm{Var}\left( \|n^{-1/2}\zeta_m^{-1}\sum_{i=1}^n\hat{R}^{m}_i\hat{\bm{U}}_i\|^2 \right)=O(1)$. Accordingly
\begin{equation*}
    \left|-n^{-1/2}\zeta_{m}^{-1}\bm{u}^\top\hat{\mathbf{D}}^{-1/2}\sum_{i=1}^n\hat{R}_i^m\hat{\bm{U}}_i\right|\le \left\|\hat{\mathbf{D}}^{-1/2}\bm{u}\right\|\left\|n^{-1/2}\zeta_m^{-1}\sum_{i=1}^n\hat{R}^{m}_i\hat{\bm{U}}_i\right\|=O_p(1).
\end{equation*}
Define $\mathbf{A}=n^{-1}\sum_{i=1}^n\hat{R}_i^{m-1}\hat{\bm{U}}_i\hat{\bm{U}}_i^\top$. After some tedious calculation, we can obtain that $\mathbb{E}(\mathrm{tr}(\mathbf{A}^2))=O(\zeta_{m-1}^2n^{-1})$. Then $\mathbb{E}\left( \bm{u}^\top\hat{\mathbf{D}}^{-1/2}\mathbf{A}\hat{\mathbf{D}}^{-1/2}\bm{u} \right)^2\le \mathbb{E}\left( \left( \bm{u}^\top\hat{\mathbf{D}}^{-1/2}\bm{u} \right)^2\mathrm{tr}\left( \mathbf{A}^2 \right) \right)=O(\zeta^2_{m-1}n^{-1})$, which leads to $\bm{u}^\top\hat{\mathbf{D}}^{-1/2}\mathbf{A}\hat{\mathbf{D}}^{-1/2}\bm{u}=O_p(\zeta_{m-1}n^{-1/2})$. Thus we have
\begin{align*}
    &p^{1/2}\zeta_m^{-1}\bm{u}\hat{\mathbf{D}}^{-1/2}\left(  \frac{1}{n}\sum_{i=1}^n\left( \hat{R}_i^{m-1}\mathbf{I}_p+(m-1)\hat{R}_i^{m-1}\hat{\bm{U}}_i\hat{\bm{U}}_i^\top  \right)\right)\hat{\mathbf{D}}^{-1/2}\bm{u}\\
    =&p^{1/2}\zeta_m^{-1} \frac{1}{n}\sum_{i=1}^n \hat{R}_i^{m-1}\bm{u}\hat{\mathbf{D}}^{-1/2} \mathbf{I}_p\hat{\mathbf{D}}^{-1/2}\bm{u}+O_p(\zeta_{m-1}n^{-1/2}),
\end{align*}
where we use the fact that $n^{-1}\sum_{i=1}^n\hat{R}^{m-1}=\zeta_{m-1}+O_p(\zeta_{m-1}n^{-1/2})$. By choosing a sufficient large $C$, the second term in \ref{eq1} dominates the first term uniformly in $\|\bm{u}\|=C$. Hence, \ref{eq1} holds and accordingly $\hat{\bm{\mu}}=O_p(b_{n,p})$
Then we prove $\hat{\bm{\mu}}=O_p(b_{n,p})$ holds as well with the same $b_{n,p}$, when $m=-1$

Consider the expansion of $\ln\left\|\hat{\mathbf{D}}^{-1/2}(\mathbf{X}_{i}-\boldsymbol{\theta}-b_{n,p}\bm{u})\right\|$:
\begin{align*}
    &\ln\left\|\hat{\mathbf{D}}^{-1/2}(\mathbf{X}_{i}-\boldsymbol{\theta}-b_{n,p}\bm{u})\right\|\\
    =&\ln\left\|\hat{\mathbf{D}}^{-1/2}(\mathbf{X}_{i}-\boldsymbol{\theta})\right\|+\frac{1}{2}\ln\left(1-2b_{n,p}\hat{R}_i^{-1}\bm{u}^\top\hat{\mathbf{D}}^{-1/2}\hat{\bm{U}_i}+b_{n,p}^2\hat{R}_i^{-2}\bm{u}^\top\hat{\mathbf{D}}^{-1}\bm{u}\right)\\
    =&\ln\hat{R}_i-b_{n,p}\hat{R}_i^{-1}\bm{u}^\top\mathbf{\hat{D}}^{-1/2}\hat{\bm{U}}_i+\frac{1}{2}b_{n,p}^2\hat{R}^{-2}_i\bm{u}\hat{\mathbf{D}}^{-1}\bm{u}+O_p(n^{-3/2}).
\end{align*}
So, it can be easily seen
\begin{equation}\label{eq2}
\begin{aligned}
    & L(b_{n,p}\bm{u})-L(\bm{0}) \\
    =&-n^{-1/2}\bm{u}^\top p^{1/2}\hat{\mathbf{D}}^{-1/2}\sum_{i=1}^n\hat{R}_i^{-1}\hat{\bm{U}}_i+\frac{p}{2}\bm{u}\hat{\mathbf{D}}^{-1}\bm{u}\sum_{i=1}^n\hat{R}_i^{-2}+O_p(n^{-1/2}).
\end{aligned}
\end{equation}
Under assumption \ref{order}, we can conclude $\mathbb{E}\left( \|n^{-1/2}p^{1/2}\sum_{i=1}^n\hat{R}^{-1}_i\hat{\bm{U}}_i\|^2 \right)=O(1)$ and $\mathrm{Var}\left( \|n^{-1/2}p^{1/2}\sum_{i=1}^n\hat{R}^{-1}_i\hat{\bm{U}}_i\|^2 \right)=O(1)$. Accordingly
\begin{equation*}
    \left|-n^{-1/2}p^{1/2}\bm{u}^\top\hat{\mathbf{D}}^{-1/2}\sum_{i=1}^n\hat{R}_i^{-1}\hat{\bm{U}}_i\right|\le \left\|\hat{\mathbf{D}}^{-1/2}\bm{u}\right\|\left\|n^{-1/2}p^{1/2}\sum_{i=1}^n\hat{R}^{-1}_i\hat{\bm{U}}_i\right\|=O_p(1).
\end{equation*}
Notice the fact that $n^{-1}\sum_{i=1}^n\hat{R}^{-2}=\zeta_{-2}+O_p(p^{-1}n^{-1/2})$. By choosing a sufficient large $C$, the second term in \ref{eq2} dominates the first term uniformly in $\|\bm{u}\|=C$. Hence, \ref{eq2} holds and accordingly $\hat{\bm{\mu}}=O_p(b_{n,p})$

As $\bm{\theta}$ is a location parameter, we assume $\bm{\theta}=0$ without loss of generality. Then $\bm{U}_i$ can be written as $\boldsymbol{U}_i=\mathbf{D}^{-1/2}\boldsymbol{X}_i/\|\mathbf{D}^{-1/2}\boldsymbol{X}_i\|=\mathbf{D}^{-1/2}\boldsymbol{\Gamma}\boldsymbol{W}_i/\|\mathbf{D}^{-1/2}\boldsymbol{\Gamma}\boldsymbol{W}_i\|$ for $i=1,2,\cdots,n.$ The estimator $\hat{\boldsymbol{\theta}}$ satisfies $\sum_{i=1}^nU(\hat{\mathbf{D}}^{-1/2}(\boldsymbol{X}_i\boldsymbol{-\hat{\boldsymbol{\theta}}}))=0$, which is is equivalent to
\begin{equation*}
    \frac{1}{n}\sum_{i=1}^{n}\hat{R}_i^m(\hat{\boldsymbol{U}}_{i}-\hat{R}_{i}^{-1}\hat{\mathbf{D}}^{-1/2}\hat{\boldsymbol{\theta}})(1-2\hat{R}_{i}^{-1}\hat{\boldsymbol{U}}_{i}^{\top}\hat{\mathbf{D}}^{-1/2}\hat{\boldsymbol{\theta}}+\hat{R}_{i}^{-2}\hat{\boldsymbol{\theta}}^{\top}\hat{\mathbf{D}}^{-1}\hat{\boldsymbol{\theta}})^{-1/2}=0.
\end{equation*}
Based on the previous discussion, we have $\|\hat{\bm{\theta}}\|=\|\hat{\bm{\mu}}\|=O_p(p^{1/2}n^{-1/2})$. By the first-Taylor expansion, the above equation can be rewritten as:
\begin{equation*}
    n^{-1}\sum_{i=1}^{n}\hat{R}^m_i\left(\hat{\boldsymbol{U}}_{i}-\hat{R}_{i}^{-1}\hat{\mathbf{D}}^{-1/2}\hat{\boldsymbol{\theta}}\right)\left(1+\hat{R}_{i}^{-1}\hat{\boldsymbol{U}}_{i}^{\top}\hat{\mathbf{D}}^{1/2}\hat{\boldsymbol{\theta}}-2^{-1}\hat{R}_{i}^{-2}\left\|\hat{\mathbf{D}}^{-1/2}\hat{\boldsymbol{\theta}}\right\|^{2}+\delta_{1i}\right)=0,
\end{equation*}
where $\delta_{1i}=O_{p}\left\{\left(\hat{R}_{i}^{-1}\hat{\bm{U}}_{i}^{\top}\hat{\mathbf{D}}^{1/2}\hat{\boldsymbol{\theta}}-2^{-1}\hat{R}_{i}^{-2}\left\|\hat{\mathbf{D}}^{-1/2}\hat{\boldsymbol{\theta}}\right\|^{2}\right)^{2}\right\}=O_{p} (n^{-1})$, which implies
\begin{equation}\label{eq3}
    \begin{aligned}
    &\frac{1}{n}\sum_{i=1}^{n}\hat{R}^m_i(1-\frac{1}{2}\hat{R}_{i}^{-2}\hat{\boldsymbol{\theta}}^{\top}\hat{\mathbf{D}}^{-1}\hat{\boldsymbol{\theta}}+\delta_{1i})\hat{\boldsymbol{U}}_{i}+\frac{1}{n}\sum_{i=1}^{n}\hat{R}_{i}^{m-1}(\hat{\boldsymbol{U}}_{i}^{\top}\hat{\mathbf{D}}^{-1/2}\hat{\boldsymbol{\theta}})\hat{\boldsymbol{U}}_{i}\\
    =&\frac{1}{n}\sum_{i=1}^{n}(1+\delta_{1i}+\delta_{2i})\hat{R}_{i}^{m-1}\hat{\mathbf{D}}^{-1/2}\hat{\boldsymbol{\theta}},
\end{aligned}
\end{equation}
where $\delta_{2i}=O_{p}(\hat{R}_{i}^{-1}\hat{\boldsymbol{U}}_{i}^{\top}\hat{\mathbf{D}}^{1/2}\hat{\boldsymbol{\theta}}-2^{-1}\hat{R}_{i}^{-2}\left\|\hat{\mathbf{D}}^{-1/2}\hat{\boldsymbol{\theta}}\right\|^{2})=O_{p}(\delta_{1i}^{1/2})$.
Similar with \cite{C23}, by Assumption \ref{order} and Markov inequality, we have that: $\max R_i^{-2}=O_p(p^{-1}n^{1/2})$, $\max\delta_{1,i}=O_p\left( \left\|\hat{\mathbf{D}}^{-1/2}\bm{\theta}\right\|^2\max \hat{R}^{-2}_i \right)=O_p(n^{-1/2})$ and $\max \delta_{2i}=O_p(n^{-1/4})$.
Considering the second term in Equation \ref{eq3}
\begin{equation*}
    \frac{1}{n}\sum_{i=1}^{n}\hat{R}_{i}^{m-1}(\hat{\bm{U}}_{i}^{\top}\hat{\mathbf{D}}^{-1/2}\hat{\boldsymbol{\theta}})\hat{\bm{U}}_{i}=\frac{1}{n}\sum_{i=1}^{n}\hat{R}_{i}^{m-1}(\hat{\bm{U}}_{i}\hat{\bm{U}}_{i}^{\top}\hat{\mathbf{D}}^{-1/2})\hat{\boldsymbol{\theta}}=\hat{\mathbf{Q}}\hat{\mathbf{D}}^{-1/2}\hat{\boldsymbol{\theta}},
\end{equation*}
where $\hat{\mathbf{Q}}=\frac{1}{n}\sum_{i=1}^n\hat{R}^{m-1}_i\hat{\bm{U}}_i\hat{\bm{U}}_i^\top$. From Lemma \ref{Qh} we acquire
\begin{equation*}
     |\hat{\mathbf{Q}}_{jl}|\lesssim \zeta_{m-1} p^{-1}|\sigma_{jl}|+O_p\left(  n^{-1/2}p^{\frac{m-3}{2}}+ p^{\frac{m-1}{2}-7/6}+ p^{\frac{m-3}{2}-\delta/2}+p^{\frac{m-1}{2}}n^{-1/2}(\log p)^{1/2}(p^{-5/2}+p^{-1-\delta/2} )\right),
\end{equation*}
and this implies that,
\begin{equation}
\begin{aligned}
    &\left\|\hat{\mathbf{Q}}\hat{\mathbf{D}}^{-1/2}\hat{\bm{\theta}}\right\|_\infty\\
    \le&\left\|\hat{\mathbf{Q}}\right\|_1\left\|\hat{\mathbf{D}}^{-1/2}\hat{\bm{\theta}}\right\|_\infty\\
    \lesssim&\zeta_{m-1}p^{-1}\|\mathbf{R}\|_1\|\hat{\mathbf{D}}^{-1/2}\hat{\bm{\theta}}\|_\infty\\
    +&O_p\left(  n^{-1/2}p^{\frac{m-3}{2}}+ p^{\frac{m-1}{2}-7/6}+ p^{\frac{m-3}{2}-\delta/2}+p^{\frac{m-1}{2}}n^{-1/2}(\log p)^{1/2}(p^{-5/2}+p^{-1-\delta/2} )\right)\|\hat{\mathbf{D}}^{-1/2}\hat{\bm{\theta}}\|_\infty\\
\end{aligned}
\end{equation}
According to Lemma \ref{bound of wU}, we obtain
\begin{align*}
    &\left\|\zeta_{m-1}^{-1}n^{-1}\sum_{i=1}^{n}\hat{R}_{i}^{m-2}\|\hat{\mathbf{D}}^{-1/2}\hat{\boldsymbol{\theta}}\|^{2}\hat{\boldsymbol{U}}_{i}\right\|_{\infty}\leq|1+H_{u}|\cdot\left\|\zeta_{m-1}^{-1}n^{-1}\sum_{i=1}^{n}\hat{R}_{i}^{m-2}\|\hat{\mathbf{D}}^{-1/2}\hat{\boldsymbol{\theta}}\|^{2}\boldsymbol{U}_{i}\right\|_{\infty}\\\lesssim& O_{p}(n^{-1})(1+O_{p}(n^{-3/2}(\log p)^{1/2}))=O_{p}(n^{-1}).
\end{align*}
Using the fact that $\zeta_{m-1}^{-1}n^{-1}\sum_{i=1}^n R_i^{m-1}=1+O_p(n^{-1/2})$ and Equation \ref{H}, we have
\begin{align*}
    &\frac{1}{n}\zeta_{m-1}^{-1}\sum_{i=1}^{n}\hat{R}_{i}^{m-1}\\=&\frac{1}{n}\zeta_{m-1}^{-1}\sum_{i=1}^{n}R_{i}^{m-1}(1+O_{p}(n^{-1/2}(\log p)^{1/2}))\\=&\left\{1+O_{p}(n^{-1/2})\right\}\left\{1+O_{p}(n^{-1/2}(\log p)^{1/2})\right\}\\=&1+O_{p}(n^{-1/2}(\log p)^{1/2}).
\end{align*}
We final obtain:
\begin{equation*}
\begin{aligned}\left\|\hat{\mathbf{D}}^{-1/2}\hat{\boldsymbol{\theta}}\right\|_{\infty}&\lesssim\left\|\zeta_{m-1}^{-1}n^{-1}\sum_{i=1}^{n}w(\hat{R}_i)\hat{\bm{U}}_{i}\right\|_{\infty}+\zeta_{m-1}^{-1}\left\|\mathbf{Q}\hat{\mathbf{D}}^{-1/2}\hat{\bm\theta}\right\|_{\infty}\\&\lesssim p^{-1}a_{0}(p)\left\|\hat{\mathbf{D}}^{-1/2}\hat{\boldsymbol{\theta}}\right\|_{\infty}+O_{p}\left\{n^{-1/2}\log^{1/2}(np)\right\}\\&+O_{p}\left(n^{-1/2}+p^{-(1/6\wedge\delta/2)}+n^{-1/2}(\log p)^{1/2}p^{-3/2}\right)\left\|\hat{\mathbf{D}}^{-1/2}\hat{\boldsymbol{\theta}}\right\|_{\infty}.\end{aligned}
\end{equation*}
Thus we conclude that:
$$\left\|\hat{\mathbf{D}}^{-1/2}\hat{\boldsymbol{\theta}}\right\|_{\infty}=O_{p}(n^{-1/2}\log^{1/2}(np)),$$
as $a_0(p)\asymp p^{1-\delta}.$
In addition, we have
$$\left\|\zeta_{m-1}^{-1}\mathbf{Q}\hat{\mathbf{D}}^{-1/2}\hat{\boldsymbol{\theta}}\right\|_{\infty}=O_{p}\left(n^{-1}\log^{1/2}(np)+n^{-1/2}p^{-(1/6\wedge\delta/2)}\log^{1/2}(np)+n^{-1}(\log p)^{1/2}p^{-3/2}\log^{1/2}(np)\right),$$
and
$$\begin{aligned}&n^{-1}\sum_{i=1}^{n}\hat{R}_{i}^{m-1}\left(1+\delta_{1i}+\delta_{2i}\right)\\=&\zeta_{m-1}\left\{1+O_{p}\left(n^{-1/4}\right)\right\}\left\{1+O_{p}(n^{-1/2}(\log p)^{1/2}\right\}\\=&\zeta_{m-1}\left\{1+O_{p}(n^{-1/4}+n^{-1/2}(\log p)^{1/2})\right\}.\end{aligned}$$
Finally, we can write
$$n^{1/2}\hat{\mathbf{D}}^{-1/2}(\hat{\boldsymbol{\theta}}-\boldsymbol{\theta})=n^{-1/2}\zeta_{m-1}^{-1}\sum_{i=1}^n\boldsymbol{U}_i+C_n,$$
where
$$\begin{aligned}\|C_{n}\|_{\infty}=&O_{p}(n^{-1/2}\log^{1/2}(np)+p^{-(1/6\wedge\delta/2)}\log^{1/2}(np)+n^{-1/2}(\log p)^{1/2}p^{-3/2}\log^{1/2}(np))\\&+O_{p}(n^{-1/4}\log^{1/2}(np)+n^{-1/2}(\log p)^{1/2}\log^{1/2}(np))\\
=&O_{p}(n^{-1/4}\log^{1/2}(np)+p^{-(1/6\wedge\delta/2)}\log^{1/2}(np)+n^{-1/2}(\log p)^{1/2}\log^{1/2}(np)).\end{aligned}$$
\end{proof}
\subsubsection{Proof of Theorem \ref{GA2}}
\begin{proof}
$$\begin{aligned}\widetilde{\rho}_{n}&=\sup_{t\in\mathbb{R}}\left|\mathbb{P}\left(n^{1/2}\left\|\hat{\mathbf{D}}^{-1/2}(\hat{\boldsymbol{\theta}}-\boldsymbol{\theta})\right\|_{\infty}\leqslant t\right)-\mathbb{P}\left(\|\boldsymbol{Z}\|_{\infty}\leqslant t\right)\right|\\&\leq\sup_{t\in\mathbb{R}}\left|\mathbb{P}\left(n^{1/2}\left\|\hat{\mathbf{D}}^{-1/2}(\hat{\boldsymbol{\theta}}-\boldsymbol{\theta})\right\|_{\infty}\leqslant t\right)-\mathbb{P}\left(\|\boldsymbol{G}\|_{\infty}\leqslant t\right)\right|+\sup_{t\in\mathbb{R}}\left|\mathbb{P}\left(\|\boldsymbol{G}\|_{\infty}\leqslant t\right)-\mathbb{P}\left(\|\boldsymbol{Z}\|_{\infty}\leqslant t\right)\right|\\&\to0.\end{aligned}$$
The last step holds from Lemma 1 and 2.
\end{proof}
\subsubsection{Proof of Theorem \ref{distribution of statistics}}
\begin{proof}
     According to the Theorem 2 in \cite{F22a}, we have
$$\mathbb{P}(p\zeta_{1}^{2}\max_{1\leq i\leq p}Z_{i}^{2}-2\log p+\log\log p\leq x)\to F(x)=\exp\left\{-\frac{1}{\sqrt{\pi}}e^{-x/2}\right\},$$
a cdf of the Gumbel distribution, as $p\to\infty.$ Thus, according to Lemma \ref{covergence of zeta} and Theorem 2
$$\begin{aligned}&|\mathbb{P}(T_{MAX}^{(m)}\leq x)-F(x)|=\left|\mathbb{P}(\hat{\zeta}_{m-1}^{2}\hat{\zeta}_{2m}^{-1}\left\|\hat{\mathbf{D}}^{1/2}(\hat{\bm{\theta}}-\bm{\theta})\right\|^2_\infty p-2\log p+\log\log p\leq x)-F(x)\right|\\
\leq&\left|\mathbb{P}(\zeta_{m-1}^{2}\zeta_{2m}^{-1}\left\|\hat{\mathbf{D}}^{1/2}(\hat{\bm{\theta}}-\bm{\theta})\right\|^2_\infty p-2\log p+\log\log p\leq x)-F(x)\right|+o(1)\\
\leq &\Big|\mathbb{P}(\zeta_{m-1}^{2}\zeta_{2m}^{-1}\left\|\hat{\mathbf{D}}^{1/2}(\hat{\bm{\theta}}-\bm{\theta})\right\|^2_\infty p-2\log p+\log\log p\leq x)-\mathbb{P}(p\zeta_{1}^{2}\max_{1\leq i\leq p}Z_{i}^{2}-2\log p+\log\log p\leq x)\Big|\\
&+\left|\mathbb{P}(p\zeta_{1}^{2}\max_{1\leq i\leq p}Z_{i}^{2}-2\log p+\log\log p\leq x)-F(x)\right|+o(1)\to0,\end{aligned}$$
for any $x\in\mathbb{R}.$
\end{proof}
\subsubsection{Proof of Theorem \ref{thpower}}
\begin{proof}
     define $T=T_{MAX}^{1/2}=n^{1/2}\|\hat{\mathbf{D}}^{-1/2}\hat{\boldsymbol{\theta}}\|_{\infty}\hat{\zeta}_{m-1}\hat{\zeta}_{2m}^{-1/2}p^{1/2}(1-n^{-1/2})^{1/2}$ and $T^c=n^{1/2}\|\hat{\mathbf{D}}^{-1/2}(\hat{\boldsymbol{\theta}}-\boldsymbol{\theta})\|_{\infty}\cdot\hat{\zeta}_{m-1}\hat{\zeta}_{2m}^{-1/2}p^{1/2}(1-n^{-1/2})^{1/2}$, which has the same distribution of $T$ under $H_0$. It is clear that$,T\geq n^{1/2}\|\hat{\mathbf{D}}^{-1/2}\boldsymbol{\theta}\|_\infty\cdot\hat{\zeta}_{m-1}\hat{\zeta}_{2m}^{-1/2}p^{1/2}(1-n^{-1/2})^{1/2}-T^c.$ Combined with Assumption and Lemma 6 we get
$$\begin{aligned}&\mathbb{P}(T_{MAX}\geq q_{1-\alpha}\mid H_{1})\\&\geq\mathbb{P}(n^{1/2}\|\hat{\mathbf{D}}^{-1/2}\boldsymbol{\theta}\|_{\infty} \hat{\zeta}_{m-1}\hat{\zeta}_{2m}^{-1/2}p^{1/2}(1-n^{-1/2})^{1/2}-T^{c}\geq{q}_{1-\alpha}\vee 0\mid H_{1})\\&=\mathbb{P}(T^{c}\leq n^{1/2}\|\hat{\mathbf{D}}^{-1/2}\boldsymbol{\theta}\|_{\infty} \hat{\zeta}_{m-1}\hat{\zeta}_{2m}^{-1/2}p^{1/2}(1-n^{-1/2})^{1/2}-{q}_{1-\alpha}\vee 0\mid H_{1})\\&\geq\mathbb{P}(T^{c}\leq n^{1/2}\left(\|\mathbf{D}^{-1/2}\boldsymbol{\theta}\|_{\infty}-\|(\mathbf{\hat{D}}^{-1/2}-\mathbf{D}^{-1/2})\boldsymbol{\theta}\|_{\infty}\right) \hat{\zeta}_{m-1}\hat{\zeta}_{2m}^{-1/2}p^{1/2}(1-n^{-1/2})^{1/2}-{q}_{1-\alpha}\vee 0\mid H_{1})\\&\geq\mathbb{P}(T^{c}\leq n^{1/2}\|\mathbf{D}^{-1/2}\boldsymbol{\theta}\|_{\infty} (1+O_{p}(n^{1/2}\log^{1/2}(np))) \hat{\zeta}_{m-1}\hat{\zeta}_{2m}^{-1/2}p^{1/2}(1-n^{-1/2})^{1/2}-{q}_{1-\alpha}\vee 0\mid H_{1})\to1,\end{aligned}$$
if$\|\boldsymbol\theta\|_\infty\geq\tilde{C}n^{-1/2}\{\log p-2\log\log(1-\alpha)^{-1}\}^{1/2}$for some large enough constant $\widetilde C.$
The last inequality holds since
$$\begin{aligned}\|(\hat{\mathbf{D}}^{-1/2}-\mathbf{D}^{-1/2})\theta\|_{\infty}&=\max_{i=1,2,\cdots,p}\frac{\hat{d}_{i}-d_{i}}{\hat{d}_{i}d_{i}}\theta_{i}\leq\max_{i=1,2,\cdots,p}|1-\frac{d_{i}}{\hat{d}_{i}}|\cdot\|\mathbf{D}^{-1/2}\boldsymbol{\theta}\|_{\infty}\\&\leq O_{p}(n^{-1/2}\log^{1/2}(np))\|\mathbf{D}^{-1/2}\boldsymbol{\theta}\|_{\infty}.\end{aligned}$$
\end{proof}
\subsubsection{Proof of Theorem \ref{powersum}}
    Theorem 5 is the special case of Theorem 6 with $\theta=0$, so we only need to show that Theorem 6 holds.
\begin{proof}
The following proof is based on the idea of the proof in article \cite{FS}, with modifications in some equations. We follow the equations in \cite{FS} on $U(\widetilde{\mathbf{D}}_{ij}^{-1/2}\boldsymbol{X}_{i})^{\top}U(\widetilde{\mathbf{D}}_{ij}^{-1/2}\boldsymbol{X}_{i}).$ By the definition, we have
\begin{align*}
&\frac{2}{n(n-1)}\sum_{1\leq i<j\leq n}R_i^mR_j^mU\left(\widetilde{\mathbf{D}}_{ij}^{-1/2}\boldsymbol{X}_{i}\right)^{\top}U\left(\widetilde{\mathbf{D}}_{ij}^{-1/2}\boldsymbol{X}_{j}\right) \\
&=\frac{2}{n(n-1)}\sum_{1\leq i<j\leq n}R_i^mR_j^m\left(\boldsymbol{U}_{i}+R_{i}^{-1}\mathbf{D}^{-1/2}\boldsymbol{\theta}+\left(\widetilde{\mathbf{D}}_{ij}^{-1/2}\mathbf{D}^{1/2}-\mathbf{I}_{p}\right)\boldsymbol{U}_{i}\right)^{\top} \\
&\times\left(\boldsymbol{U}_{j}+R_{j}^{-1}\mathbf{D}^{-1/2}\boldsymbol{\theta}+\left(\widetilde{\mathbf{D}}_{ij}^{-1/2}\mathbf{D}^{1/2}-\mathbf{I}_{p}\right)\boldsymbol{U}_{j}\right)(1+\alpha_{ij})^{-1/2}\left(1+\alpha_{ji}\right)^{-1/2} \\
&=\frac{2}{n(n-1)}\sum_{1\leq i<j\leq n}R_i^mR_j^m\boldsymbol{U}_{i}^{\top}\boldsymbol{U}_{j}+\frac{2}{n(n-1)}\sum_{1\leq i<j\leq n}R_{i}^{m-1}R_{j}^{m-1}\boldsymbol{\theta}^{\top}\mathbf{D}^{-1}\boldsymbol{\theta} \\
&+\frac{2}{n(n-1)}\sum_{1\leq i<j\leq n}R_i^mR_j^m\boldsymbol{U}_{i}^{\top}\boldsymbol{U}_{j}\left[(1+\alpha_{ij})^{-1/2}\left(1+\alpha_{ji}\right)^{-1/2}-1\right] \\
&+\frac{4}{n(n-1)}\sum_{1\leq i<j\leq n}R_i^mR_j^mU_{i}^{\top}\left(\widetilde{\mathbf{D}}_{ij}^{-1/2}\mathbf{D}^{1/2}-\mathbf{I}_{p}\right)\boldsymbol{U}_{j}\left(1+\alpha_{ij}\right)^{-1/2}\left(1+\alpha_{ji}\right)^{-1/2} \\
&+\frac{2}{n(n-1)}\sum_{1\leq i<j\leq n}R_i^mR_j^m\boldsymbol{U}_{i}^{\top}\left(\widetilde{\mathbf{D}}_{ij}^{-1/2}\mathbf{D}^{1/2}-\mathbf{I}_{p}\right)^{2}\boldsymbol{U}_{j}\left(1+\alpha_{ij}\right)^{-1/2}\left(1+\alpha_{ji}\right)^{-1/2} \\
&+\frac{2}{n(n-1)}\sum_{1\leq i<j\leq n}R_i^mR_{j}^{m-1}\boldsymbol{\theta}^{\top}\mathbf{D}^{-1}\boldsymbol{\theta}\left[\left(1+\alpha_{ij}\right)^{-1/2}\left(1+\alpha_{ji}\right)^{-1/2}-1\right] \\
&+\frac{4}{n(n-1)}\sum_{1\leq i<j\leq n}R_i^mR_j^mU_{i}^{\top}\mathrm{D}^{-1/2}\theta\left(1+\alpha_{ij}\right)^{-1/2}\left(1+\alpha_{ji}\right)^{-1/2} \\
&+ \frac{2}{n(n-1)}\sum_{1\leq i<j\leq n}R_i^mR_j^m\boldsymbol{U}_{i}^{\top}\left(\widetilde{\mathbf{D}}_{ij}^{-1/2}\mathbf{D}^{1/2}-\mathbf{I}_{p}\right)\mathbf{D}^{-1/2}\boldsymbol{\theta}\left(1+\alpha_{ij}\right)^{-1/2}\left(1+\alpha_{ji}\right)^{-1/2} \\
&:= \frac{2}{n(n-1)}\sum_{1\leq i<j\leq n}R_i^mR_j^m\boldsymbol{U}_{i}^{\top}\boldsymbol{U}_{j}+\frac{2}{n(n-1)}\sum_{1\leq i<j\leq n}R_{i}^{m-1}R_{j}^{m-1}\boldsymbol{\theta}^{\top}\mathbf{D}^{-1}\boldsymbol{\theta} \\
&+A_{n1}+A_{n2}+A_{n3}+A_{n4}+A_{n5}+A_{n6}.
\end{align*}
where
$$\alpha_{ij}=2R_i^{-1}\boldsymbol{U}_i^\top\left(\widetilde{\mathbf{D}}_{ij}^{-1/2}-\mathbf{D}^{-1/2}\right)\boldsymbol{X}_i+R_i^{-2}\left\|\left(\widetilde{\mathbf{D}}_{ij}^{-1/2}-\mathbf{D}^{-1/2}\right)\boldsymbol{X}_i\right\|^2+2R_i^{-1}\mathbf{D}^{-1/2}\boldsymbol{\theta}+R_i^{-2}\boldsymbol{\theta}^\top\mathbf{D}^{-1}\boldsymbol{\theta}.$$
Note that $R_i^{- 1}\bm{U}_i^\top \left ( \widetilde{\mathbf{D} } _{ij}^{- 1/ 2}- \mathbf{D} ^{- 1/ 2}\right ) \boldsymbol{X}_i= \boldsymbol{U}_i^\top \left ( \widetilde{\mathbf{D} } _{ij}^{- 1/ 2}\mathbf{D} ^{1/ 2}- \mathbf{I} _p\right ) \boldsymbol{U}_i+ R_i^{- 1}\boldsymbol{U}_i^\top \left ( \widetilde{\mathbf{D} } _{ij}^{- 1/ 2}- \mathbf{D} ^{- 1/ 2}\right ) \boldsymbol{\theta }=$
$O_{p}\left(n^{-1/2}\left(\log p\right)^{1/2}\right)$ and $R_i^{-2}\left\|\left(\widetilde{\mathbf{D}}_{ij}^{-1/2}-\mathbf{D}^{-1/2}\right)\boldsymbol{X}_{i}\right\|^{2}=O_{p}\left(n^{-1}\log p\right)$by Lemma \ref{Convergence of D} By Assumption \ref{R2} and Equation \ref{ha} $R_{i}^{- 1}\mathbf{D} ^{- 1/ 2}\boldsymbol{\theta }= O_{p}\left ( n^{- 1}\right )$ and $R_i^{-2}\boldsymbol{\theta}^{\top}\mathbf{D}^{-1}\boldsymbol{\theta}=O_{p}\left(n^{-2}\right)$ So $\alpha_{ij}=O_{p}\left(n^{-1/2}\left(\log p\right)^{1/2}\right).$

Similarly, we will show that $A_{n1}=o_p(\sigma_n)$ where $\sigma_n^2=2n^{-2}p^{-2}\zeta_{2m}\mathrm{tr}(\mathbf{R}^2)$. Under some calculations, we get $\mathbb{E}[(R_i^mR_j^m\boldsymbol{U}_i^\top\boldsymbol{U}_j)^2]=\mathrm{tr}(\bm{\Sigma}_w^2)$By Lemma \ref{Qh} we find that $\Sigma_{w,i,j}=\zeta_{2m}p^{-1}\sigma_{i,j}+O(p^{m-1-\delta/2}).$ Thus we have
$$\begin{aligned}\mathrm{tr}(\Sigma_{w}^{2})&=\sum_{i=1}^{p}\sum_{j=1}^{p}\Sigma_{w,i,j}^{2}=\sum_{i=1}^{p}\sum_{j=1}^{p}\left(\zeta_{2m}^2p^{-2}\sigma_{i,j}^{2}+\sigma_{i,j}O(p^{2m-2-\delta/2})\right)\\&=\sum_{i=1}^{p}\sum_{j=1}^{p}\zeta_{2m}^2p^{-2}\sigma_{i,j}^{2}+\sum_{p^{-\delta/2}=O(\sigma_{i,j})}\sigma_{i,j}O(p^{2m-2-\delta/2})\\&+\sum_{\sigma_{ij}\in[C_{1}\frac{O(t(\mathbf{R}^{2}))}{p^{2}-V^{2}},C_{2}p^{-\theta/2}]}\sigma_{i,j}O(p^{2m-2-\delta/2})+\sum_{\sigma_{ij}=O(\frac{O(t(\mathbf{R}^{2}))}{p^{2}-V^{2}})}\sigma_{i,j}O(p^{2m-2-\delta/2})\\&=\zeta_{2m}^2p^{-2}\mathrm{tr}(\mathbf{R}^{2})(1+O(1))+O(p^{2m-2-\delta/2})\cdot\frac{p^{2-\delta/2}}{O(\mathrm{tr}(\mathbf{R}^{2}))}\cdot o(\frac{p^{2}}{n})+p^{-2}O(\mathrm{tr}(\mathbf{R}^{2}))\\&=O(p^{2m-2}\mathrm{tr}(\mathbf{R}^{2})).\end{aligned}$$
By the Cauchy inequality,
$$\begin{aligned}E\left(A_{n1}^{2}\right)&=O\left(n^{-4}\right)\sum_{i<j}E\left\{R_i^mR_j^m\boldsymbol{U}_{i}^{T}\boldsymbol{U}_{j}\left[\left(1+\alpha_{ij}\right)^{-1/2}\left(1+\alpha_{ji}\right)^{-1/2}-1\right]\right\}^{2}\\
&\leq O\left(n^{-2}\right)E\left(R_i^mR_j^m\boldsymbol{U}_{i}^{T}\boldsymbol{U}_{j}\right)^{2}E\left[\left(1+\alpha_{ij}\right)^{-1/2}\left(1+\alpha_{ji}\right)^{-1/2}-1\right]^{2}\\
&=O\left(n^{-3}\log p\left(p^{2m-2}\mathrm{tr}(\mathbf{R}^{2})+O(p^{2m-2-\delta/2})(p+no(\frac{p}{n^{1/2}}))\right)\right)=o\left(\sigma_{n}^{2}\right).\end{aligned}$$
$$\begin{aligned}A_{n2}&=\frac{4}{n(n-1)}\sum_{i<j}R_i^mR_j^m\bm{U}_{i}^{\top}\left(\widetilde{\mathbf{D}}_{ij}^{-1/2}\mathbf{D}^{1/2}-\mathbf{I}_{p}\right)\boldsymbol{U}_{j}\\&+\frac{4}{n(n-1)}\sum_{i<j}R_i^mR_j^m\bm{U}_{i}^{\top}\left(\widetilde{\mathbf{D}}_{ij}^{-1/2}\mathbf{D}^{1/2}-\mathbf{I}_{p}\right)\boldsymbol{U}_{j}\left[\left(1+\alpha_{ij}\right)^{-1/2}\left(1+\alpha_{ji}\right)^{-1/2}-1\right]\\
&:=G_{n1}+G_{n2}.
\end{aligned}$$
By Lemma \ref{Convergence of D} and Equation ${\mathbb{E}}((R_i^mR_j^m\boldsymbol{U}_{i}^{\top}\left(\widetilde{\mathbf{D}}_{ij}^{-1/2}\mathbf{D}^{1/2}-\mathbf{I}_{p}\right)\boldsymbol{U}_{j})^{2})\leq O(n^{-1}\log p$ tr$(\boldsymbol\Sigma_{w}^{2}))=$ $o(p^{2m-2}$tr$(\mathbf{R}^2)).$Then we obtain $G_{n1}=o_p(\sigma_n).$ Similar to $A_{n1}$,we can show $G_{n2}=o_p(\sigma_n)$ Taking the same procedure as $A_{n2}$, we can obtain $A_{n3}=o_p(\sigma_n).$ Similarly to the processing of Equation \ref{111} we get
$$\frac{2}{n(n-1)}\sum_{1\leq i<j\leq n}R_i^m R_j^m\bm{U}_{i}^{\top}\bm{U}_{j}=\frac{2}{n(n-1)}\sum_{1\leq i<j\leq n}R_i^m R_j^m(\mathbf{D}^{-1/2}\Gamma U(W_{i}))^{\top}\mathbf{D}^{-1/2}\Gamma U(W_{j})+o_{p}(\sigma_{n}).$$
Similar to proof in \cite{FS}, by Lemma \ref{Qh} we final acquire
$$\frac{2}{\sigma_n n
(n-1)}\sum_{1\leq i<j\leq n}R_i^m R_j^m(\mathbf{D}^{-1/2}\Gamma U(\boldsymbol{W}_{i}))^{\mathsf{T}}\mathbf{D}^{-1/2}\Gamma U(\boldsymbol{W}_{j})\stackrel{d}{\to}N(0,1).$$
\end{proof}
\subsubsection{Proof of Theorem \ref{a.i.1}}
The following proof is similar to the proof of in \cite{L24}. The main difference from it is that we replace $\bm{U}_i$ by $R_i^m\bm{U}_i$.
\begin{proof}
    To prove $T_{SUM}^{(m)}$ and $T_{MAX}^{(m)}$ are asymptotically independent, it suffices to show that, under $H_0$,
$$\mathbb{P}(\dfrac{T_{SUM}^{(m)}}{\sigma_n}\leq x,T_{MAX}^{(m)}\leq y)\to\Phi(x)\cdot\exp\left\{-\dfrac{1}{\sqrt{\pi}}e^{-y/2}\right\}.$$
From the proof of Theorem 2 in \cite{FS},we acquire
$$T_{SUM}^{(m)}=\frac{2}{n(n-1)}\sum\sum_{i<j}R_i^{m}R_j^{m}\boldsymbol{U}_i^\top\boldsymbol{U}_j+o_p(\sigma_n),$$
and it's easy to find that $\sigma_{n}^{2}=\frac{2\zeta_{2m}^2}{n(n-1)p}+o(\frac{1}{n^{3}})$ according to Assumption \ref{R1} ,Combined with Theorem \ref{Bahadur represengtation} we can get
$$\begin{aligned}&\mathbb{P}(\frac{\frac{2}{n(n-1)}\sum\sum_{i<j}R_i^mR_j^m\boldsymbol{U}_{i}^{\top}\boldsymbol{U}_{j}}{\sigma_{n}}+o_{p}(1)\leq x\\&,p\left\|n^{-1/2}\zeta_{2m}^{-1/2}\sum_{i=1}^{n}R_i^m\boldsymbol{U}_{i}\right\|_{\infty}^{2}+O_{p}(L_{n,p})\leq y)\\&\to\Phi(x)\cdot\exp\left\{-\frac{1}{\sqrt{\pi}}e^{-y/2}\right\}.\end{aligned}$$
We next prove that,
\begin{equation}\label{123}
\begin{aligned}
&\mathbb{P}(\sqrt{\frac{n}{n-1}}\left(\frac{\|\sqrt{\frac{p}{n\zeta_{2m}}}\sum_{i=1}^{n}R_i^m\boldsymbol{U}_{i}\|_{2}^{2}-p}{\sqrt{2\mathrm{tr}(\mathbf{R}^{2})}}\right)\leq x,\left\|\sqrt{\frac{p}{n\zeta_{2m}}}\sum_{i=1}^{n}R_i^m\boldsymbol{U}_{i}\right\|_{\infty}^{2}\leq y)\\&\to\Phi(x)\cdot\exp\left\{-\frac{1}{\sqrt{\pi}}e^{-y/2}\right\}.
\end{aligned}
\end{equation}

When Equation holds, combined with $O_p(L_{n,p})=o_p(1)$, Equationholds obviously,
which means that the independence of $T_{SUM}^{(m)}$ and $T_{MAX}^{(m)}$ follows.

From the Theorem 2 in \cite{F22a}, the Equation \ref{123} holds if $\boldsymbol{U}_i$ follows the normal distribution. We then investigate the non-normal case. Let $\boldsymbol{\xi}_i=$ $\boldsymbol{U}_i\in R^p,i=1,2,\cdots,n.$ For $\boldsymbol{z}=(z_1,\cdots,z_q)^\top\in R^q$, we consider a smooth approximation of the maximum function, namely,
$$F_\beta(z):=\beta^{-1}\log(\sum_{j=1}^q\exp(\beta z_j)),$$
where $\beta>0$ is the smoothing parameter that controls the level of approximation. An elementary calculation shows that for all $z\in R^q$,
$$0\leq F_{\beta}(\boldsymbol{z})-\max_{1\leq j\leq q}z_{j}\leq\beta^{-1}\log q.$$
Define $\sigma_{S}^{2}=2n^{2}$tr$(R^{2})$,
$$\begin{aligned}W(\boldsymbol{x}_{1},\cdots,\boldsymbol{x}_{n})&=\frac{\|\sqrt{\frac{p}{n}}\sum_{i=1}^{n}\boldsymbol{x}_{i}\|_{2}^{2}-p}{\sqrt{2\mathrm{tr}(\mathbf{R}^{2})}}\\&=\frac{p\sum_{i\neq j}\boldsymbol{x}_{i}^{\top}\boldsymbol{x}_{j}}{\sqrt{2n^{2}}\mathrm{tr}\:(\mathbf{R}^{2})}:=\frac{p\sum_{i\neq j}\boldsymbol{x}_{i}^{\top}\boldsymbol{x}_{j}}{\sigma_{S}},\\V(\boldsymbol{x}_{1},\cdots,\boldsymbol{x}_{n})&=\beta^{-1}\log(\sum_{j=1}^{p}\exp(\beta\sqrt{\frac{p}{n}}\sum_{i=1}^{n}x_{i,j})).\end{aligned}$$
By setting $\beta=n^{1/8}\log(n)$, Equation \ref{123} is equivalent to
$$P(W(\boldsymbol{\xi}_{1},\cdots,\boldsymbol{\xi}_{p})\leq x,V(\boldsymbol{\xi}_{1},\cdots,\boldsymbol{\xi}_{p})\leq u_{p}(y))\to\Phi(x)\cdot\exp(-\exp(y)).$$
Suppose $\{\boldsymbol Y_1,\boldsymbol{Y}_2,\cdots,\boldsymbol{Y}_n\}$ are sample from $N(0,\zeta_{2m}^{-1}\mathbb{E}R_i^{2m}\boldsymbol{U}_1^{\top}\boldsymbol{U}_1)$, and independent with $\zeta_{2m}^{-1/2}\boldsymbol{U}_1,\cdots,\zeta_{2m}^{-1/2}\boldsymbol{U}_n$ or write as $\boldsymbol{\xi}_1,\cdots,\boldsymbol{\xi}_n).$ The key idea is to show that: $(W(\boldsymbol{\xi}_1,\cdots,\boldsymbol{\xi}_n),V(\boldsymbol{\xi}_1,\cdots,\boldsymbol{\xi}_n))$ has the same limiting distribution as ${W}(\boldsymbol{Y}_1,\cdots,\boldsymbol{Y}_n),V(\boldsymbol{Y}_1,\cdots,\boldsymbol{Y}_n)).$
The following proof is almost the same as the proof in Theorem 6 in \cite{L24}, except that $\bm{\Sigma}_u$ is replaced by $\zeta_{2m}^{-1}\bm{\Sigma}_w$
\end{proof}
\subsubsection{Proof of Theorem \ref{a.i.2}}
\begin{proof}
    From the proof of Theorem 2 in \cite{FS},we can find that
$$T_{SUM}=\frac{2}{n(n-1)}\sum\sum_{i<j}R_i^m R_j^m\bm{U}_{i}^{\top}\bm{U}_{j}+\zeta_{m-1}^{2}\boldsymbol{\theta}^{\top}\mathbf{D}^{-1}\boldsymbol{\theta}+o_{p}(\sigma_{n}),$$
and according to Lemma ll with minor modifications, we get the Bahadur representation in $L^{\infty}$ norm,
$$n^{1/2}\mathbf{D}^{-1/2}(\hat{\boldsymbol{\theta}}-\boldsymbol{\theta})=n^{-1/2}\zeta_{m-1}^{-1}\sum_{i=1}^{n}(R_i^m\boldsymbol{U}_{i}+\zeta_{m-1}\mathbf{D}^{-1/2}\boldsymbol{\theta})+C_{n}.$$
Similar with the proof in Theorem 6 it's suffce to show the result holds for normal
version, i.e. it suffce to show that:
$$\|\sqrt{\dfrac{p}{n}}\sum_{i=1}^n\boldsymbol{Y}_i\|^2\text{and}\|\sqrt{\dfrac{p}{n}}\sum_{i=1}^n(\boldsymbol{Y}_i+\zeta_{2m}^{-1/2}\zeta_{m-1}\mathbf{D}^{-1/2}\boldsymbol{\theta})\|_\infty^2,$$
are asymptotic independent, where $\{\bm{Y}_1,\bm{Y}_2,\cdots,\bm{Y}_n\}$ are sample from $N(0,\zeta_{2m}^{-1}\mathbb{E}R_i^{2m}\boldsymbol{U}_1^\top\boldsymbol{U}_1).$
Denote $\sqrt{\frac pn}\sum_{i=1}^{n}\bm{Y}_{i}:=\bm{\varphi}=(\varphi_{1},\cdots,\varphi_{p})^{\top},\bm{\varphi}_{\mathcal{A}}=(\varphi_{j_{1}},\cdots,\varphi_{j_{d}})^{\mathsf{T}}$,and $\varphi_{\mathcal{A}^{c}}=(\varphi_{j_{d+1}},\cdots,\varphi_{j_{p}})^{\mathsf{T}}$,

where $\mathcal{A}=\{j_{1},j_{2},\cdots,j_{d}\}.$ Then$,S=\|\bm{\varphi}\|^{2}=\|\bm{\varphi}_{A}\|^{2}+\|\bm{\varphi}_{A^{c}}\|^{2},M=\|\bm{\varphi}+\sqrt{np}\zeta_{2m}^{-1/2}\zeta_{m-1}\mathbf{D}^{-1/2}\boldsymbol{\theta}\|_{\infty}=\operatorname*{max}_{i\in\mathcal{A}}(\varphi_{i}+\sqrt{np}\zeta_{2m}^{-1/2}\zeta_{m-1}\mathbf{D}^{-1/2}\boldsymbol{\theta})+\operatorname*{max}_{i\in\mathcal{A}^{c}}\varphi_{i}.$From the proof of Theorem \ref{a.i.1} we know that$\|\boldsymbol\varphi_{\mathcal{A}^{c}}\|^{2}$ and $\max_{i\in\mathcal{A}^c}\varphi_i$ are asymptotically independent. Hence, it suffce to show that $\|\boldsymbol\varphi_{\mathcal{A}^c}\|^2$ is asymptotically independent with $\varphi_{\mathcal{A}}.$

By Lemma \ref{14} $\varphi_{A^{c}}$ can be decomposed as $\varphi_{A^{c}}=E+F$, where $E=\varphi_{A^{c}}-\Sigma_{U,A^{c},A}\Sigma_{U,A,A}^{-1}\varphi_{A},F=\Sigma _{U, \mathcal{A} ^{c}, \mathcal{A} }\Sigma _{U, \mathcal{A} , \mathcal{A} }^{- 1}\varphi _{\mathcal{A} }, \Sigma _{U}= p\zeta_{2m}^{-1}\mathbb{E} R_i^{2m}\boldsymbol{U}_{1}\boldsymbol{U}_{1}^{\top }= p\zeta_{2m}^{-1}\Sigma _{w}$ , which fulfill the properties $\boldsymbol E\sim N(0,\boldsymbol\Sigma_{U,\mathcal{A}^{c},\mathcal{A}^{c}}-\Sigma_{U,\mathcal{A}^{c},\mathcal{A}}\Sigma_{U,\mathcal{A},\mathcal{A}}^{-1}\Sigma_{U,\mathcal{A},\mathcal{A}^{c}}),\boldsymbol{F}\sim N(0,\Sigma_{U,\mathcal{A}^{c},\mathcal{A}}\Sigma_{U,\mathcal{A},\mathcal{A}^{-1}}\Sigma_{U,\mathcal{A},\mathcal{A}^{c}})$ and $\boldsymbol E$ and $\boldsymbol\varphi_{\mathcal{A}}$ are independent.
Then, we rewrite
$$\|\varphi_{A^{c}}\|^{2}=E^{\top}E+F^{\top}F+2E^{\top}F.$$
According the proof of lemma S.7 in \cite{F22a}, we have that:
$$\mathbb{P}(|\boldsymbol{F}^\top\boldsymbol{F}+2\boldsymbol{E}^\top\boldsymbol{F}|\geq\epsilon\nu_p)\leq\frac{3}{p^t}\to0,$$
by $d= o( \lambda _{\mathrm{min}}( \mathbf{R} )$tr$(\mathbf{R}^{2})^{1/2}/(\log p)^{C})$ ,where $\nu_p=[2$tr$( \mathbf{R} ^{2}) ] ^{1/ 2}, t= t_{p}: = C\epsilon / 8\cdot v_{p}/ [ \lambda _{\mathrm{max}}( \mathbf{R} ) \log p] \to$
$\infty , \epsilon _{p}: = ( \log p) ^{C}/ [ v_{p}\lambda _{\mathrm{min}}( \mathbf{R} ) ] \to 0.$
\end{proof}

\end{document}